\documentclass[amsmath,pra,twocolumn]{revtex4-2}
\usepackage{times}
\usepackage{epsfig}
\usepackage{amssymb}
\usepackage{amsmath}
\usepackage{amsfonts}
\usepackage{tcolorbox} 
\usepackage{bm}
\usepackage{multirow}
\usepackage{graphicx}
\usepackage[colorlinks=true, allcolors=blue]{hyperref}

\begin{document}
\title{Effect of antiprotons on hydrogen-like ions in external magnetic fields}

\author{A. Anikin$^{1,2}$}
\email[E-mail:]{alexey.anikin.spbu@gmail.com}
\author{A. Danilov$^{1}$}
\author{D. Glazov$^{1,4}$}
\author{A. Kotov$^1$}
\author{D. Solovyev$^{1,4}$}
%\email[E-mail:]{d.solovyev@spbu.ru}
\affiliation{ 
$^1$ Department of Physics, St. Petersburg State University, Peterhof, Oulianovskaya 1, 198504, St. Petersburg, Russia\\
$^2$ D.I. Mendeleev Institute for Metrology, St. Petersburg, 190005, Russia\\
$^3$ School of Physics and Engineering, ITMO University, Kronverkskiy pr. 49, 197101 St. Petersburg, Russia\\
$^4$ Petersburg Nuclear Physics Institute named by B.P. Konstantinov of National Research Centre 'Kurchatov Institut', St. Petersburg, Gatchina, 188300, Russia
}

\begin{abstract}
%In the present work quasi-molecular compounds consisting of one antiproton ($\bar{p}$) and one hydrogen-like ion are investigated: $\mathrm{He}^{+} - \bar{p}$, $\mathrm{Li}^{2+} - \bar{p}$, $\mathrm{C}^{5+} - \bar{p}$, $\mathrm{S}^{15+} - \bar{p}$, $\mathrm{Kr}^{35+} - \bar{p}$,  $\mathrm{Ho}^{66+} - \bar{p}$, $\mathrm{Re}^{74+} - \bar{p}$, $\mathrm{U}^{91+} - \bar{p}$. Adiabatic potential curves are constructed in the framework of the A-DKB approach using the numerical solution of the Dirac equation with a two-center potential by means of the finite basis method adapted to systems with axial symmetry. The Zeeman shifts in the case of a homogeneous magnetic field along the axis of the molecule are calculated.

In the present work, quasi-molecular compounds consisting of one antiproton ($\bar{p}$) and one hydrogen-like ion are investigated: $\mathrm{He}^{+} - \bar{p}$, $\mathrm{Li}^{2+} - \bar{p}$, $\mathrm{C}^{5+} - \bar{p}$, $\mathrm{S}^{15+} - \bar{p}$, $\mathrm{Kr}^{35+} - \bar{p}$, $\mathrm{Ho}^{66+} - \bar{p}$, $\mathrm{Re}^{74+} - \bar{p}$, $\mathrm{U}^{91+} - \bar{p}$. For the calculations, the Dirac equation with two-center potential is solved numerically using the dual-kinetically balanced finite-basis-set method adapted to systems with axial symmetry (A-DKB). Adiabatic potential curves are constructed for the ground state of the above quasi-molecular compounds in the framework of the A-DKB approach. Calculations were also performed for the case of an external magnetic field (the field is taken into account non-perturbatively). Zeeman shifts of the quasi-molecular terms are obtained for a homogeneous magnetic field with a strength of the laboratory order (up to 100 Tesla) directed along the axis of the molecule.
\end{abstract}

\maketitle

\section{Introduction}

Investigation of hydrogen-like ions (ions with only one electron) is of great interest in many fields of physics. Throughout the whole periodic table H-like ions serve as one of a key tools to measure electronic $g$-factor \cite{verdu2004determination, beier2003measurement, vogel2005towards, 10.1063/1.57483, Kohler_2015, PhysRevLett.96.253002}, Lamb shift \cite{https://doi.org/10.1002/andp.201800324, 10.1063/1.3292459, Gassner_2018, Kraft-Bermuth_2017} etc. Of particular importance is the investigation of hyperfine structure of the one-electron highly-charged ions (HCI). Being one of the most sensitive tools for testing quantum electrodynamics (QED), such systems represent the basis for studies of the structure of the nucleus, approaching the sector of strong interactions %(which uses its own perturbative computational methods, such as the $1/Z$ decomposition on the nucleus charge) %Being one of the most sensitive tools to test quantum electrodynamics (QED), such systems provide a somewhat link between the nuclear structure studies and atomic physics
\cite{PhysRevA.57.879,PhysRevLett.73.2425}. In turn, collisions of HCI with both neutral atoms and other ions, leading to the formation of quasi-molecular systems, provide information for studying critical QED phenomena of bound states, such as the creation of electron-positron pairs \cite{gershtein1970positron, pieper1969interior, Ya_B_Zeldovich_1972, RAFELSKI1978227, shabaev2019qed, PhysRevLett.123.113401, PhysRevD.102.076005, universe7040104}. Quasi-molecules consisting of H-like HCIs have been widely studied in literature, see, for example, \cite{kotov2020ground, kotov2021one, kotov2022} and references therein. 

The study of antiprotons colliding with atoms, molecules and hydrogen-like ions is another interesting subject of modern physics \cite{doi:10.1098/rsta.2017.0271}. Light antiproton systems offer an opportunity for the successful study of ionisation processes \cite{kirchner2011current} or the formation of protonium atoms ($p-\bar{p}$) \cite{sakimoto2004protonium}. Matter-antimatter compounds can also be used to test fundamental interactions or search for effects that violate known physical symmetries: such systems are used to straightforward verification of CPT-invariance by comparing the spectra of matter and antimatter \cite{PhysRevA.98.010101}. The collisions of heavy HCI and atoms with antiprotons can be also used to study the antiproton capture processes \cite{genkin2009possibility} and such phenomenon as Coulomb glory \cite{demkov1984new, Yu_N_Demkov_2001, PhysRevA.76.032709}.

%Being a difficult task for exact theoretical solution, the many-body problem arising in the study of quasimolecules is solved by numerical calculations.
Referring to the well-known problem of solving the many-body problem exactly, the study of quasi-molecules is carried out using numerical methods. One of the most widely used and effective approaches is the nonrelativistic theory. Within this theory the one-electron problem is solved with high accuracy, whereas relativistic and quantum electrodynamics (QED) effects are described in the framework of nonrelativistic quantum electrodynamics (NRQED) \cite{korobov2007relativistic1,tsogbayar2006relativistic,korobov2020hyperfine}. While providing a powerful tool for calculations in the case of light systems, NRQED is a poor approximation for heavy element compounds. In this case, it is more advantageous to solve the one-electron problem for dinuclear compounds using the Dirac equation with the corresponding two-center Coulomb potential. As in NRQED, this method uses the Born-Oppenheimer approximation along with various decompositions of the two-center potential \cite{PhysRevA.86.052705,tupitsyn2014relativistic,mironova2015relativistic}.

In this paper, we apply a dual-kinetic-balance (DKB) method based on the decomposition of the wave function over the finite basis set of B-splines \cite{Shabaev_DKB}, generalized to the axially symmetric systems (A-DKB) \cite{Rozenbaum_ADKB}. The DKB method is widely used in various relativistic and QED calculations for highly charged ions. Recently, the A-DKB method has been successfully worked out to describe both heavy \cite{kotov2020ground, kotov2021one, kotov2022} and light \cite{anikin2023light, danilov2023, Solovyev_2024} few-electron quasi-molecules. The present work is devoted to the study of quasi-molecules consisting of antiproton and different (light and heavy) hydrogen-like ions subjected to an external uniform magnetic field directed along the molecular axis. %Light (quasi)molecular systems are studied in very strong magnetic fields, up to $10^{11} - 10^{12}$ T, as in the atmospheres of neutron stars, see \cite{wille1988magnetically, turbiner2004h, olivares2010one}, and also \cite{schmelcher1997molecules, lehtola2020fully}.
In this study, we restrict ourselves to fields with strengths relevant to laboratory conditions. Experimental research on HCI is often carried out using ion traps. The external magnetic field in these devices is typically a few Tesla \cite{beiersdorfer1999x}, whereas to date, experimentally achievable magnetic field strengths have reached the extraordinary value of $100$ T (National High Magnetic Field Laboratory \cite{kasahara2017upper, modic2017robust}). It should be noted that the Born-Oppenheimer approximation used in the present work is valid only as a rough approach in the case of very strong magnetic fields. For the fields of the order of up to $10^{11} - 10^{12}$ T, as in the atmospheres of neutron stars \cite{wille1988magnetically, turbiner2004h,olivares2010one,schmelcher1997molecules,lehtola2020fully}, the contribution of the nuclei motion is comparable to the Coulomb binding energy and requires special consideration \cite{schmelcher1988electronic}. %In this article, we restrict ourselves to the case of laboratory fields.

The paper is organized as follows. A brief description of the method used is presented in section~\ref{ADKB}. Results of the electronic energy calculations are given in section~\ref{antip}, in which nuclear charge dependence of the adiabatic potentials is analyzed, as well as dependence of Zeeman shifts on both nuclear charge and on the inter-nuclear distance. The last section provides a discussion of the obtained numerical results. Numerical results are collected in Tables~\ref{tab:1}-\ref{tab:9}. The atomic units $\hbar=e=m_{e}=1$ ($\hbar$ is the Planck constant, $e$ is the electron charge modulus, $m_{e}$ is the mass of electron) are used throughout the paper. In these units, $c$ (the speed of light) can be expressed through the fine structure constant $c=1/\alpha$, but for clarity it is written explicitly.

\section{The dual kinetic balance method for systems with axial symmetry (A-DKB)}
\label{ADKB}
	
\begin{figure}[ht]
\centering
\includegraphics[width=0.8\columnwidth]{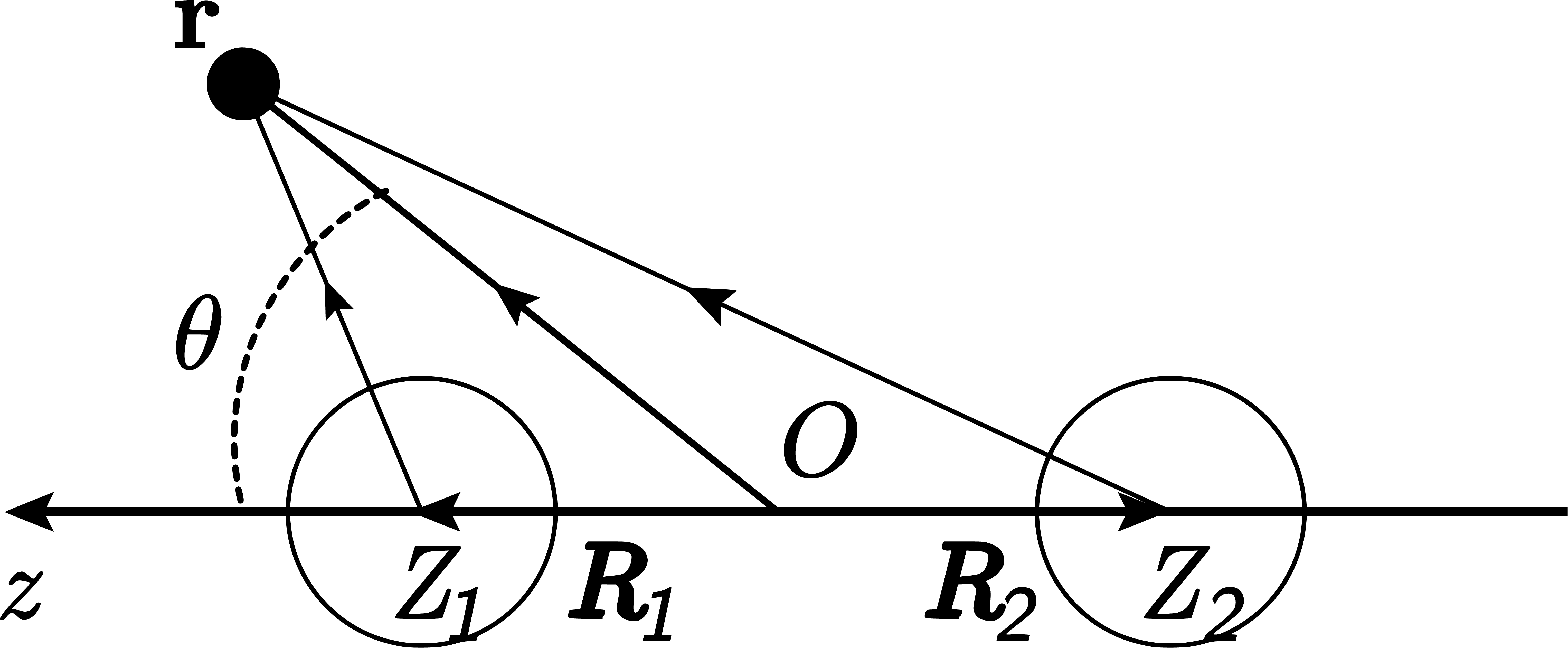}
\caption{The coordinate system for diatomic one-electron compounds oriented along the $z$-axis. The bound electron is marked with a filled circle, the nuclei are marked with circles with the nucleus charges $Z_1$ and $Z_2$. The corresponding position vectors from the coordinate origin $O$ are indicated by $\boldsymbol{R}_1$ and $\boldsymbol{R}_2$. The position vector of the bound electron placed on one of the nuclei is given by $\boldsymbol{r}$ directed from $O$. Since axial symmetry, the azimuth angle $\theta$ is introduced. }
\label{fig0}
\end{figure}

In frames of the Born-Oppenheimer approximation the stationary Dirac equation for the two-center potential is
\begin{eqnarray}
\label{Dirac}
%\left[ \boldsymbol{\alpha}\boldsymbol{p} + \beta + V(\boldsymbol{r}) \right] \Psi_{n}(\boldsymbol{r}) = E_n \Psi_{n} (\boldsymbol{r}),
\left[ c\boldsymbol{\alpha}\boldsymbol{p} + \beta c^2+ V(\boldsymbol{r}) \right] \Psi_{n}(\boldsymbol{r}) = E_n \Psi_{n} (\boldsymbol{r}),
\end{eqnarray} 
where $\boldsymbol{r}$ is the electron radius vector, $\boldsymbol{p}$ is the momentum operator and $\boldsymbol{\alpha}$, $\beta$ are the Dirac matrices. The two-center potential is expressed by
\begin{eqnarray}
\label{pot}
V(\boldsymbol{r}) = V_1(|\boldsymbol{r} - \boldsymbol{R}_1|) + V_2(|\boldsymbol{r} - \boldsymbol{R}_2|)
\\
\nonumber
%\equiv -\frac{\alpha Z_1}{\left|\bm{r}-\bm{R}_1\right|}-\frac{\alpha Z_2}{\left|\bm{r}-\bm{R}_2\right|},
\equiv -\frac{Z_1}{\left|\bm{r}-\bm{R}_1\right|}-\frac{Z_2}{\left|\bm{r}-\bm{R}_2\right|},
\end{eqnarray}
with the Coulomb potentials $V_{1,2}$, generated by nuclei with charges $Z_1$ and $Z_2$ at the point $\boldsymbol{r}$ and located at a distance of $R\equiv|\bm{R}|=|\bm{R}_1-\bm{R}_2|$ from each other. While in Eq. (\ref{pot}) the potential is written for point-like charges, in the A-DKB calculations various finite-nucleus charge models are used. In the present work we nonperturbatevely study effect of the external magnetic field directed along the molecular axis. This magnetic interaction is accounted for by the operator 
\begin{eqnarray}
    \label{magn}
    V_{\mathrm{ext}} = -\frac{e}{2} ([\boldsymbol{r} \times \boldsymbol{\alpha}] \cdot \boldsymbol{B}),
\end{eqnarray}
where $e/2$ is Bohr magneton (in a.u.) and $\boldsymbol{B}$ is magnetic field strength. The coordinate origin is chosen between the two nuclei, the $z$ axis coincides with the molecule axis, see Fig.~\ref{fig0}. 

Within the relativistic approach, the wave function can be written as
\begin{eqnarray}
\label{wf}
\psi\left(r, \theta, \varphi \right) = \frac{1}{r}
\begin{pmatrix}
G_1\left( r, \theta \right) e^{i(m_j - 1/2)\varphi}  \\
G_2\left( r, \theta \right) e^{i(m_j + 1/2)\varphi} \\
i F_1\left( r, \theta \right) e^{i(m_j - 1/2)\varphi} \\
i F_2\left( r, \theta \right) e^{i(m_j + 1/2)\varphi}
\end{pmatrix},
\end{eqnarray}
where $m_j$ is the projection of the total angular momentum of the electron onto the nuclear axis, which is conserved in the axially symmetric case. Then, the components $G_i(r,\theta)$ and $F_i(r,\theta)$ ($i=1,2$), according to the A-DKB method \cite{Rozenbaum_ADKB}, are decomposed through $B$-splines as follows:
\begin{eqnarray}
\label{bsplinesexpan}
\phi \left( r, \theta \right) 
\cong \sum^{4}_{u = 1} \sum^{N_r}_{i_r = 1} \sum^{N_{\theta}}_{i_{\theta} = 1} C^{u}_{i_r i_{\theta}} \Lambda B_{i_r} (r) Q_{i_{\theta}} (\theta) e_u.
\end{eqnarray}
Here $\left\{ B_{i_r} (r) \right\}^{N_r}_{i_r = 1}$ are the B-splines, $\left\{ Q_{i_{\theta}} (\theta) \right\}^{N_{\theta}}_{i_{\theta} = 1}$ are given by the Legendre polynomials of the argument $2\theta/\pi - 1$, and $\left\{ e_u \right\}^4_{u = 1}$ are the four-component basis vectors. To avoid the problem of spurious states \cite{Johnson_Bspline,Sapirstein_1996}, the matrix $\Lambda$ introduced in Eq. (\ref{bsplinesexpan}) imposes the dual-kinetic-balance (DKB) conditions on the basis set \cite{Shabaev_DKB,Rozenbaum_ADKB}. The eigenvalues and eigenfunctions are found numerically by solving the generalized eigenvalue problem. 

Application of the DKB approach for axially symmetric systems is implied for the extended charge nucleus only. The point-like nucleus case can be accessed by the extrapolation of the extended-nucleus results in vanishing nuclear size. The expression above is given for arbitrary basis sets $\{B_{i_r}(r)\}_{i_r=1}^{N_r}$ and $\{Q_{i_\theta}(r)\}_{i_\theta=1}^{N_\theta}$. A particular choice of one-component basis functions corresponds to $B$-splines of the second order and Legendre polynomials as $Q_{i_\theta}(\theta) = P_{i_\theta-1}\left(\frac{2\theta}{\pi}-1\right)$, forming the set of one-component $\theta$-dependent basis functions of polynomial degrees $l=0\dots N_\theta-1$. By varying the values $N_r$, $N_\theta$, such a sample allows the convergence of numerical calculations to be monitored. 

In the following we consider the ground states of one-electron quasi-molecules consisting of one hydrogen-like ion and one antiproton: $\mathrm{He}^{+} - \bar{p}$, $\mathrm{Li}^{2+} - \bar{p}$, $\mathrm{C}^{5+} - \bar{p}$, $\mathrm{S}^{15+} - \bar {p}$, $\mathrm{Kr}^{35+} - \bar{p}$, $\mathrm{Ho}^{66+} - \bar{p}$, $\mathrm{Re}^ {74+} - \bar{p}$, $\mathrm{U}^{91+} - \bar{p}$. This choice covers a wide range of $Z_1$, permitting the main features to be identified, and the corresponding H-like ions have been studied theoretically and experimentally, see e.g. \cite{beiersdorfer1999x, lopez1996direct, lopez1998nuclear,seelig1998ground, georgiadis1986measurement}. The magnetic field strength considered in this paper is: $1-5$, $10$ and $100$ Tesla.

% \section{Effect of antiprotons on hydrogen like ions}
\section{Results}
\label{antip}

To study the antiproton impact on hydrogen-like ions, we calculate fully relativistic electronic energies within the A-DKB approach. The following grid parameters were used: $N_r = 395$ and $N_{\theta} = 40$, see Eq.~(\ref{bsplinesexpan}). For the H-like ions the nuclei radii were taken from \cite{angeli2013table} and for the antiproton $r_{\bar{p}} = r_p = 0.8410$ fm was used ($r_p$ is the proton charge radius \cite{RevModPhys.93.025010}). We used the spherical nucleus model for light ions $\mathrm{He}^{+}$, $\mathrm{Li}^{2+}$, $\mathrm{C}^{5+}$, and the antiproton, and the Fermi model for heavy ion nuclei $\mathrm{S}^{15+}$, $\mathrm{Kr}^{35+}$, $\mathrm{Ho}^{66+}$, $\mathrm{Re}^{74+}$, $\mathrm{U}^{91+}$. To study compounds with such a broad range of nuclear charges, for every element we consider a "chemical" distance, which in atomic units equals to $2/Z_1$. For binuclear quasi-molecules, this distance is often used as a characteristic scale, and in this paper, for a given $Z_1$, the electronic energy is calculated for inter-nuclear distances within the interval $R \in [ 0; 4/Z_1]$ (twenty one points from zero to twice the chemical distance in a.u.). Recently, the A-DKB method has been used in \cite{anikin2023light,danilov2023,Solovyev_2024,kotov2020ground,kotov2021one,kotov2022} to study both light and heavy one-electron binuclear compounds. Accuracy was achieved at $10^{-8}$ relative magnitude for the grid parameters given above. For the Zeeman shifts the same level of accuracy was reached in the present work for the considered systems. % Therefore, in the this work we expect such level of accuracy and present the results with corresponding number of digits.

\begin{figure}[ht!]
\centering
\caption{Adiabatic potential energy curves for the ground state of H-like ion-antiproton compounds. The inter-nuclear distance is scaled as $\tilde{R} \rightarrow R \times Z_1$, and the energy is depicted according to $\tilde{E}_{\mathrm{A-DKB}} \rightarrow E_{\mathrm{A-DKB}} / Z_1^{2 }$. All values are given in a.u.}
\includegraphics[width=1\linewidth]{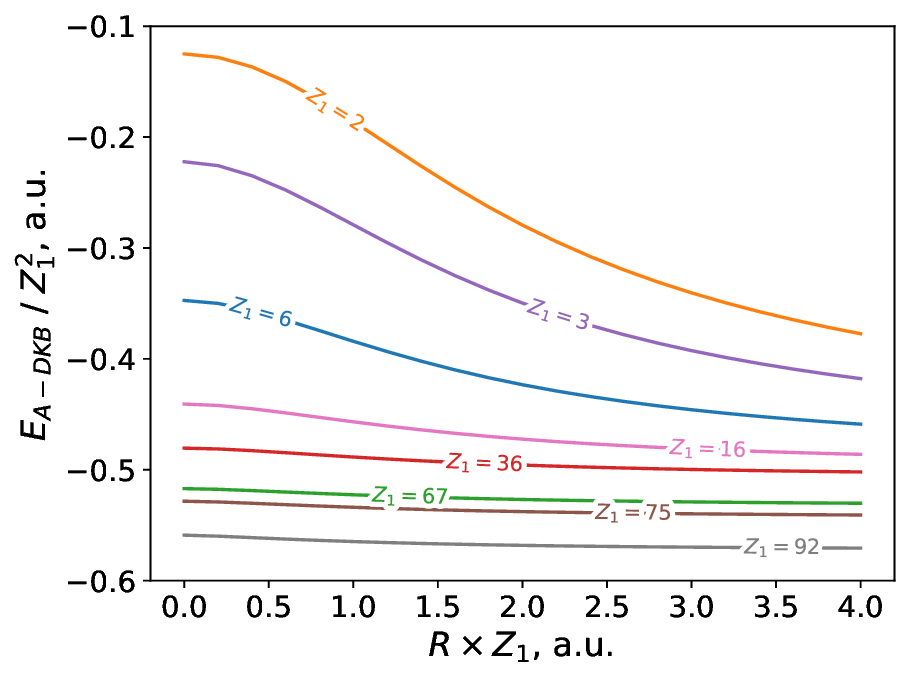}
\label{fig1}
\end{figure}

%First of all, consider the ground states adiabatic potential energy curves of the H-like ions bounded with antiproton. For this purpose, using the results of the A-DKB calculations listed after the main text in Tables~\ref{tab:2}-\ref{tab:9} we scaled the internuclear distances as $R \rightarrow R \times Z$ and the energies as $E_{\mathrm{A-DKB}} \rightarrow E_{\mathrm{A-DKB}} / Z^{2}$. Corresponding graphs are plotted on Fig.~\ref{fig1}.
Adiabatic electron energy curves scaled according to the relation $E_{\mathrm{A-DKB}} \rightarrow E_{\mathrm{A-DKB}} / Z_1^{2}$, for inter-nuclear distances $\tilde{R} \rightarrow R \times Z_1$ are represented by the plots in Fig.~\ref{fig1}. The numerical values for each compound can be found in Tables~\ref{tab:2}-\ref{tab:9}, given for brevity at the end of the paper.

Fig.~\ref{fig1} shows that the greater the nuclear charge, the weaker the influence of the antiproton on the corresponding ion. In particular, an almost threefold energy change is evident for the quasi-molecule $\mathrm{He}^{+} - \bar{p}$ (upper curve), whereas for the compound $\mathrm{U} ^{91+} - \bar{p}$ the electronic energy variation is about $\approx 10 \%$ (lower curve). This behaviour is expected for the following reasons. When the charge of the nucleus is large, the electron is exposed to a Coulomb field that is orders of magnitude (especially for $\mathrm{U}^{91+}$) larger than that of the antiproton, and hence the bound electron almost does not "feel" the antiproton. The situation is quite different for the light elements: %the charges $Z = 2$ and $Z = -1$ strongly influence each other, and the energy of the electron.
the total field of charges $Z_1 = 2$ and $Z_2 = -1$ considerably weakens with the approach of the antiproton to the one-electron ion, which affects the electronic binding energy. Among this, ionisation occurs for $Z_1=1$ and $Z_2 = -1$, see e.g. \cite{anikin2023light} and references therein. 

From the graphs in Fig~\ref{fig1}, it also follows that for ions with higher nucleus charge, relativistic corrections combined with a correction for finite nucleus size have to be taken into account. In particular, the chosen scaling allows one to clearly see the difference of the obtained electron energies from their nonrelativistic values. In the limit of the joined atom ($R \rightarrow 0$) or infinitely distant nuclei ($R \rightarrow \infty$), the nonrelativistic energy for a point nucleus is $E_{\mathrm{nr}}^{(R=0)} = -(Z_1-1)^2 / 2$ and $E_{\mathrm{nr}}^{(R=\infty)} = -Z_1^2 / 2$ in atomic units. The latter gives $0.5$ a.u. when using the above scaling, whereas the former is approximately $0.49$ for $\mathrm{U} ^{91+} - \bar{p}$ compound.

%In particular, one can see that, for example in case of hydrogen-like uranium the curve is significantly far from $0.5$ a.u. If the relativistic effects and the correction for the finite size of the nucleus were negligibly small, this curve would be close to the $0.5$ a.u. In the limit of the joined ($R \rightarrow 0$) as well as of infinitely separated ($R \rightarrow \infty$) nuclei nonrelativistic energy for point-like nucleus is $E_{\mathrm{nr}} = -Z^2 / 2$ a.u. (with $Z$ corresponding to the limit considered), which under the scaling used on Fig.~\ref{fig1} will give $0.5$ a.u. For heavy elements significant discrepancy of the electronic energy from $0.5$ a.u. shows the necessity of taking into account relativistic effects and finite nuclear size corrections, both are accounted for in the A-DKB calculations.

% \subsection{Influence of external magnetic field}
% \label{antip_field}

The main goal of the present work, however, is to study the influence of antiprotons on H-like ions in an external homogeneous magnetic field. In the framework of the A-DKB approach, such a combination can be analysed quite simply for a field directed along the molecule axis. The calculations can be performed in a non-perturbative way, i.e., by including the interaction of Eq. (\ref{magn}) in the Hamiltonian of Eq. (\ref{Dirac}). %To this end, we nonperturbatevely calculated the fully relativistic electronic energies in the magnetic fields for different values of the field strength ($1-5, 10, 100$ T) and evaluated the corresponding Zeeman shifts, $\Delta E_{\mathrm{Z}}$.
To this end, the fully relativistic electron energies have been calculated in magnetic fields with different values of field strength ($1-5$, $10$, $100$ Tesla). Then the corresponding Zeeman shifts $\Delta E_{\mathrm{Z }}$ can be retrieved by comparing with numerical results obtained for the case of absence in the Hamiltonian of interaction with the external field. The results are combined in Table~\ref{tab:1} for chemical distances, and at the end of the paper Tables~\ref{tab:2}-\ref{tab:9} present the values for inter-nuclear distances in the range $R \in [0; 4/Z]$ a.u. All Zeeman shifts coincide in sign with the sign of the projection of the electron total angular momentum onto the molecular axis.
\begin{table}[ht]
\centering
\caption{Absolute values of Zeeman shifts, $\Delta E_{\mathrm{Z}}$, in atomic units (a.u.) for compounds of hydrogen-like ions with an antiproton. The ions considered are listed in the first column, the magnetic field strength values in Tesla (T) in the second row of the table. All values in the table are calculated for chemical inter-nuclear distances ($ 2/Z $ a.u.) and multiplied by a factor $10^6$.}
\label{tab:1}
\resizebox{\columnwidth}{!}{
\begin{tabular}{ c | c c c c c c c }
\hline
\hline
\multicolumn{1}{c|}{\multirow{2}{*}{ion}} & \multicolumn{7}{c}{$| \Delta E_{\mathrm{Z}} |$ $\times 10^6$, a.u. (at a chemical distance, 2/Z a.u.)} \\ \cline{2-8} 
\multicolumn{1}{c|}{} & $B = 1$ T     & $B = 2$ T     & $B = 3$ T     & $B = 4$ T     & $B = 5$ T     & $B = 10$ T     & $B = 100$ T    \\ \hline
$\mathrm{He}^{+}$     & $ 2.13 $      & $ 4.25 $      & $ 6.38 $      & $ 8.51 $      & $ 10.63 $     & $ 21.27 $      & $ 212.69 $     \\
$\mathrm{Li}^{2+}$    & $ 2.13 $      & $ 4.25 $      & $ 6.38 $      & $ 8.51 $      & $ 10.63 $     & $ 21.27 $      & $ 212.68 $     \\
$\mathrm{C}^{5+}$     & $ 2.13 $      & $ 4.25 $      & $ 6.38 $      & $ 8.50 $      & $ 10.63 $     & $ 21.26 $      & $ 212.58 $     \\
$\mathrm{S}^{15+}$    & $ 2.12 $      & $ 4.23 $      & $ 6.35 $      & $ 8.47 $      & $ 10.59 $     & $ 21.17 $      & $ 211.76 $     \\
$\mathrm{Kr}^{35+}$   & $ 2.08 $      & $ 4.15 $      & $ 6.23 $      & $ 8.31 $      & $ 10.39 $     & $ 20.77 $      & $ 207.76 $     \\
$\mathrm{Ho}^{66+}$   & $ 1.95 $      & $ 3.89 $      & $ 5.84 $      & $ 7.79 $      & $ 9.73 $      & $ 19.46 $      & $ 194.65 $     \\
$\mathrm{Re}^{74+}$   & $ 1.90 $      & $ 3.79 $      & $ 5.69 $      & $ 7.59 $      & $ 9.48 $      & $ 18.96 $      & $ 189.65 $     \\
$\mathrm{U}^{91+}$    & $ 1.76 $      & $ 3.52 $      & $ 5.29 $      & $ 7.05 $      & $ 8.81 $      & $ 17.62 $      & $ 176.17 $     \\ \hline\hline
\end{tabular}
}
\end{table}

%As one finds from Tab.~\ref{tab:1}, Zeeman shifts are linear in the field strength. In case of the (weak) fields considered in the present paper, to calculate energy shifts one can use perturbation theory, in frames of which linear in perturbation behavior is rather expected. Interestingly enough is that the Zeeman shift value is almost does not depend on the nuclear charge value in case of the light elements and varies significantly for heavy elements. Again, the higher the value of the nuclear charge, the more important it is to take into account relativistic effects and finit nuclear size corrections, which are taken into account in the A-DKB calculations. From Tab.~\ref{tab:2}-\ref{tab:9} one can also analyze a dependence of the Zeeman shifts on internuclear distance. For light elements, $\Delta E_{Z}$ is practically independent of the internuclear distances: at $10, 100$ T the shift changes only at $10^{-7}-10^{-8}$ a.u. level. As the nuclear charge increases, the Zeeman shifts begin to depend weakly on the distance between the nuclei: for example, for $\mathrm{U}^{92+} - \bar{p}$ at a field of $100$ T, the Zeeman shift changes by $\sim 10^{-6}$ a.u. as the distance changes from $R = 0$ fm to $R \approx 2000$ fm.

As follows from Table~\ref{tab:1}, the Zeeman shifts are linear in the field strength for all compounds considered. In the case of (relatively weak) fields involved in this study, perturbation theory can be used to calculate the energy shifts, in which a linear approximation is sufficient. It can be found that the value of the Zeeman shift is almost independent of the nuclear charge in the case of light elements and varies significantly for heavy ions (starting from $Z_1\sim 36$ in Table~\ref{tab:1}).

From Tables~\ref{tab:2}-\ref{tab:9} one can also analyze a dependence of the Zeeman shifts on inter-nuclear distance. For light elements, $\Delta E_{Z}$ is practically independent of the inter-nuclear distance: at $10$, $100$ T the shift varies only at the level of $10^{-7}-10^{-8}$ a.u. As the nuclear charge increases, the Zeeman shift begins to depend weakly on the inter-nuclear distance. This is already apparent for the hydrogen-like Krypton Kr$^{35+}$. For the compound $\mathrm{U}^{92+} - \bar{p}$ at a field of $100$ T, the Zeeman shift changes by $\sim 10^{-6}$ a.u. as the distance changes from $R = 0$ fm to $R \approx 2000$ fm.

\section{Discussion and conclusions}

In this paper, within the framework of a completely relativistic method the effect of antiprotons on various hydrogen-like ions is studied using the following quasi-molecules as an example: $\mathrm{He}^{+} - \bar{p}$, $\mathrm{Li}^{2+} - \bar{p}$, $\mathrm{C}^{5+} - \bar{p}$, $\mathrm{S}^{15+} - \bar{p}$, $\mathrm{Kr}^{35+} - \bar{p}$,  $\mathrm{Ho}^{66+} - \bar{p}$, $\mathrm{Re}^{74+} - \bar{p}$, $\mathrm{U}^{91+} - \bar{p}$. To obtain the ground state electron energy of these compounds, we utilized the Dual Kinetic Balance Approach \cite{Shabaev_DKB}, extended to the Dirac equation for axially symmetric systems in \cite{Rozenbaum_ADKB}. The A-DKB method has already proven its effectiveness for both heavy \cite{kotov2020ground, kotov2021one, kotov2022} and light \cite{anikin2023light, danilov2023, Solovyev_2024} one-electron quasi-molecules, so in this study we focused on the electron energy of the ground term and its dependence on the nuclear charge.

To analyze how presence of antiproton affects hydrogen-like ions with different nuclear charges we calculated adiabatic potential energy curves and plotted the corresponding scaled graphs, see Fig.~\ref{fig1}. It was found that, as one might expect, for elements with higher nuclear charges relativistic effects and corrections due to the finite nuclear size play an increasingly important role. At the same time, for higher nuclear charges, due to the large Coulomb potential, the electron energy is almost independent of the inter-nuclear distance, in contrast to the light elements.

%Separate attention has been paid to study the influence of external magnetic field along the molecular axis on electronic energy of hydrogen-like ions in presence of antiproton.
The study of the antiproton effect on the energy states of a bound electron in hydrogen-like ions with different nuclear charges has been extended to the case of the presence of an external magnetic field.
For this purpose adiabatic potential energy curves has been obtained for $1-5$, $10$, $100$ T magnetic fields. As a consequence the Zeeman shifts were determined, see numerical results in Tables~\ref{tab:1} - \ref{tab:9}. 

It has been established that in the presence of an external field, light and heavy compounds behave quite differently. In particular, the Zeeman shift for an arbitrary nuclear charge can be estimated with high accuracy as linear field strength contribution in the considered range of $1-100$ T. However, its value vary significantly with $Z_1$. The effect is manifested in a weak dependence of the Zeeman shift on the charge of the nucleus. At large $Z_1$ ($Z_1\geqslant 36$), the deviation becomes noticeable and increases with growing field strength. It should be emphasized that in the nonrelativistic limit, the Zeeman shift does not depend on $Z_1$. Thus, this behavior refers to relativistic effects in combination with the finite size of the nucleus, which are taken into account in the A-DKB method.

Moreover, from Tables~\ref{tab:2} - \ref{tab:9} it is clear that in the case of small nuclear charges, Zeeman shifts are practically independent of the inter-nuclear distance (up to double the chemical distance $4/Z_1$). For $\mathrm{He}^{+} - \bar{p}$ even at $100$ T the corresponding change is of the order of $10^{-8}$ a.u. absolute value, i.e. at the level of the numerical error of the A-DKB method. For large nuclear charges $Z_1$, for example in the compound $\mathrm{U}^{92+} - \bar{p}$, such a slight change in the Zeeman shift is observed for the field $1$ T, and, hence, at $100$ T it increases to $10^{-6}$ a.u. However, even such a result can be considered independent of the inter-nuclear distance with an accuracy limit of $10^{-6}$.

Additionally, the results collected in Tables~\ref{tab:2}-\ref{tab:9} can serve to better determine the capture energy of antiprotons by hydrogen-like ions. In particular, the binding energy of the electron initial state $E_{\mathrm{A-DKB}}$, given in the second column for each specified ion in the mentioned tables, includes the entire relativistic value taking into account the effect of finite nucleus size. The latter is more pronounced for heavy ions. For example, the Dirac energy of the bound electron for hydrogen-like uranium is $E_{\mathrm D}=-4861.1979$ a.u. (the antiproton is infinitely distant) and with correction for the finite size of the nucleus amounts to $-4853.8957$ a.u. \cite{PhysRevA.102.042811}. The bound electron energy found in the framework of the A-DKB approach is $E_{\mathrm{A-DKB}}\approx -4830.4778$ a.u. at the inter-nuclear distance $R=0.043478$ a.u. or $2300.77065$ fm (see the last row in Table~\ref{tab:9}), which therefore represents a significant shift of the resonance of the antiproton capture process \cite{Genkin_2009}. At the same time, the level shift in the magnetic field, even with a strength of $100$ T, can be considered as a correction of the next order of smallness: the variation of the energy of the bound electron state differs by hundreds of eV with approaching antiproton to the nucleus and the Zeeman shift in the field of $100$ T is, respectively, meV, remaining practically unchanged. 

Finally, the correction for the finite nucleus size should be analysed in conjunction with the model used to describe it. As was recently observed in \cite{anikin2023light,Solovyev_2024}, in light binuclear compounds the electronic spectrum is independent of the nuclear size correction with an accuracy of about $10^{-8}$ for the relative magnitude. Obviously, this correction becomes significant for heavy ions. In turn, antiproton compounds should also be sensitive to this effect. Since the nuclear field changes essentially as the aniti-proton approaches, the model used to describe the nucleus is also important. To demonstrate this, the relevant calculations are given in the following Table~\ref{tab:nm} (for brevity, for the U$^{91+}-\bar{p}$ compound only).
\begin{table}[ht]
\centering
\caption{The binding energy of an electron for the ground term in a quasi-molecule U$^{91+}-\bar{p}$. Energy values are given in atomic units (a.u.) for inter-nuclear distances of $0.0$, $1150.383$ fm ($2/Z_1$), and $2300.771$ fm ($4/Z_1$) in the corresponding columns. To determine the Zeeman shift the electron energies are given in each first and second rows for the zero field and the field strength $10$ T, respectively. The Zeeman shift, $\Delta E_Z$, multiplied by a factor of $10^6$ is indicated in every third row in a.u. To reveal the energy dependence on the finite nucleus size model, the radius (in fm) and the model used for comparison are given in the first column.}
\label{tab:nm}
% зависимость от модели и от радиуса для урана
% 80-40, расстояния 0.0, 1150.383, 2300.771, радиус из Анжели 5.8571 (для данного радиуса и модели в первой строке энергия без поля, во второй с полем, а в третьей Зееман; по колонкам - расстояния)
% ------- отделены числа, которые в статье
\resizebox{\columnwidth}{!}{
\begin{tabular}{ c | c | c | c }
\hline
%radius, fm &&& \\
Fermi-model   & $R=0.0$ fm & $R=1150.383$ fm & $R=2300.771$ fm \\
%Zeeamn shift &&& \\
\hline
\hline
$r^{-}_N=5.8538$ fm & $-4731.62494651$ & $-4809.44455498$ & $-4830.48421214$ \\
$B=10$ T & $-4731.62496422$ & $-4809.44457260$ & $-4830.48422976$ \\
$\Delta E_Z\times 10^6$, a.u. & $17.71$ & $17.62$ & $17.61$ \\

\hline

$r_N=5.8571$ fm\footnote{This row contains the values given in Table~\ref{tab:9}.} & $-4731.61940826$ & $-4809.43830386$ & $-4830.47781493$ \\
$B=10$ T & $-4731.61942597$ & $-4809.43832148$ & $-4830.47783255$ \\
$\Delta E_Z\times 10^6$, a.u. & $17.71$ & $17.62$ & $17.61$ \\

\hline

$r^{+}_N=5.8604$ fm & $-4731.61386859$ & $-4809.43204318$ & $-4830.47141770$ \\
$B=10$ T & $-4731.61388630$ & $-4809.43206080$ & $-4830.47143531$ \\
$\Delta E_Z\times 10^6$, a.u. & $17.71$ & $17.62$ & $17.62$ \\

\hline
\hline
sphere-model   & $R=0.0$ fm & $R=1150.383$ fm & $R=2300.771$ fm \\
\hline
\hline
            
$r^{-}_N=5.8538$ fm & $-4731.61239329$ & $-4809.41349294$ & $-4830.41091172$ \\
$B=10$ T & $-4731.61241100$ & $-4809.41351056$ & $-4830.41092934$ \\
$\Delta E_Z\times 10^6$, a.u. & $17.71$ & $17.62$ & $17.62$ \\

\hline
            
$r_N=5.8571$ fm & $-4731.60685922$ & $-4809.40704558$ & $-4830.40439320$ \\
$B=10$ T & $-4731.60687693$ & $-4809.40706320$ & $-4830.40441081$ \\
$\Delta E_Z\times 10^6$, a.u. & $17.71$ & $17.62$ & $17.61$ \\

\hline
            
$r^{+}_N=5.8604$ fm & $-4731.60132373$ & $-4809.40059737$ & $-4830.39787536$ \\
$B=10$ T & $-4731.60134144$ & $-4809.40061499$ & $-4830.39789297$ \\
$\Delta E_Z\times 10^6$, a.u. & $17.71$ & $17.62$ & $17.62$ \\
\hline
\hline
\end{tabular}
}
\end{table}

First, the values listed in each column of Table~\ref{tab:nm} show the independence of the Zeeman shift from the radius and the nuclear model used. This follows from the fact that these effects shift the energy term equally for cases with and without a field. Thus, the Zeeman shift in our work is determined with an accuracy of $10^{-4}$ relative magnitude or $10^{-8}$ on the absolute scale for the compound U$^{91+}-\bar{p}$. %The small deviation in the numbers (at the $10^{-8}$ level) relates more to the accuracy of the numerical calculations. 

\begin{figure}[ht]
\centering
\includegraphics[width=1\columnwidth]{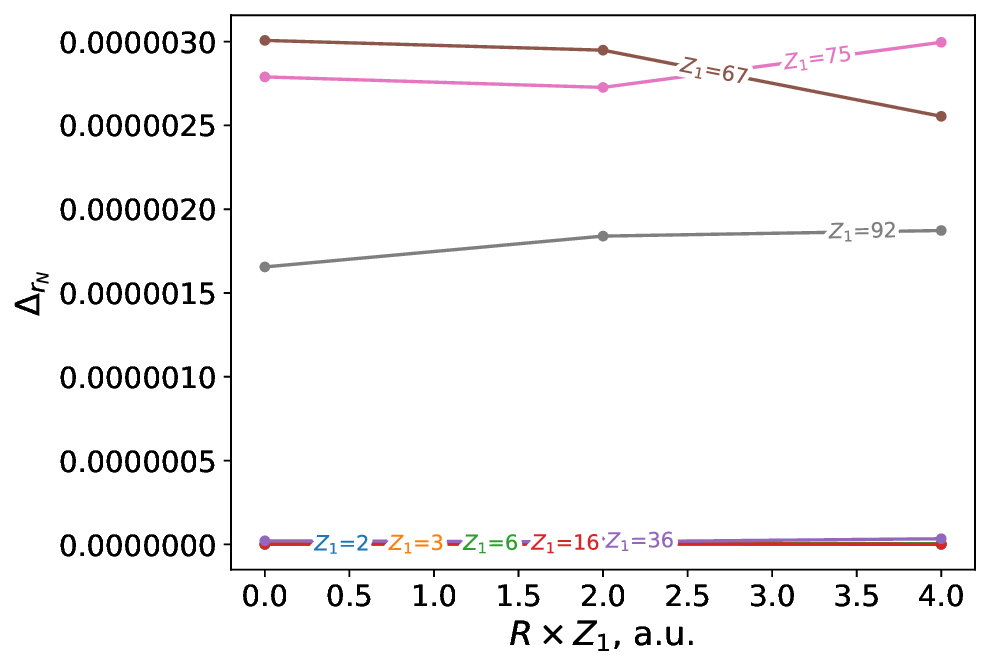}
\caption{The graph showing the dependence of the relative magnitude $\Delta_{r_{N}}$ on the inter-nuclear distance, $R$. }
\label{fig2}
\end{figure}

Second, the dependence of the results on the nuclear radius can be observed by comparing the corresponding rows in each sector of the table. In particular, one finds that the results vary by about $10^{-7}$ when changing the radius near the error of its determination \cite{angeli2013table}, i.e. $r_N=5.8571(33)$ fm. %By observing the behaviour over the inter-nuclear distance, we can estimate the effect on the error of numerical calculations. With the removal of the antiproton, the binding energies should be less sensitive to the nuclear radius of the heavy ion, since at large distances from the nucleus the field is determined purely by the Coulomb interaction (as for a point charge). Therefore, the deviation from such an intuitive estimate can be attributed to the accuracy of the calculation within the A-DKB framework.
%In the spherical model for light and Fermi model for heavy elements, this behaviour is shown as a deviation $\Delta_{r_{N}} = \left|E_{\mathrm{A-DKB}}^{( r_{N}+ )} - E_{\mathrm{A-DKB}}^{(r_{N}-) }\right|/E_{\mathrm{A-DKB}}^{(r_{N})}$ depending on the inter-nuclear distance (for $0.0$, $2 / Z_1$ and $4 / Z_1$ distances) in Fig.~\ref{fig2}, for which additional calculations similar to those listed in Table~\ref{tab:nm} were performed for elements $\mathrm{He}^{+} - \bar{p}$, $\mathrm{Li}^{2+} - \bar{p}$, $\mathrm{C}^{5+} - \bar{p}$, $\mathrm{S}^{15+} - \bar{p}$, $\mathrm{Kr}^{35+} - \bar{p}$,  $\mathrm{Ho}^{66+} - \bar{p}$, $\mathrm{Re}^{74+} - \bar{p}$, $\mathrm{U}^{91+} - \bar{p}$.
To demonstrate the dependence on the nuclear radius at different inter-nuclear distances for the elements $\mathrm{He}^{+} - \bar{p}$, $\mathrm{Li}^{2+} - \bar{p}$, $\mathrm{C}^{5+ } - \bar{p}$, $\mathrm{S}^{15+} - \bar{p}$, $\mathrm{Kr}^{35+} - \bar{p}$, $\mathrm {Ho}^{66+} - \bar{p}$, $\mathrm{Re}^{74+} - \bar{p}$, $\mathrm{U}^{91+} - \bar{p}$, plots of the relative value of $\Delta_{r_{N}} = \sqrt{ (\Delta_{r^+_{N}})^2 + (\Delta_{r^-_{N}})^2 }$, where $\Delta_{r^{\pm}_{N}} = \left|E_{\mathrm{A-DKB}}^{( r^{\pm}_{N} )} - E_{\mathrm{A-DKB}}^{(r_{N})}\right|/E_{\mathrm{A-DKB}}^{(r_{N})}$ are shown in Fig.~\ref{fig2}. The following notations are introduced for the relative deviation $\Delta_{r_{N}}$: $r_{N}$, $r^-_{N}$, and $r^+_{N}$ denote the nuclear radius, nuclear radius minus error, and nuclear radius plus error, respectively, see \cite{angeli2013table}. For the light elements ($Z_1\leq 6$) the spherical model of the nucleus is used, and for the others the Fermi model is employed.
It should be noted that the errors in the values of the nuclear radii \cite{angeli2013table} differ significantly for different nuclei. For example, $r_\mathrm{Re} = 5.3596 \pm 0.0172$, whereas $r_\mathrm{U} = 5.8571 \pm 0.0033$, which leads to a considerably stronger deviation of $\Delta_{r_{N}}$ for hydrogen-like ions with $Z_1=75$ and $Z_1=92$. The graph in Fig.~\ref{fig2} shows that in the case of light elements the dependence on the nucleus size is an order of magnitude smaller compared to heavy nuclei.

\begin{figure}[ht]
\centering
\includegraphics[width=1\columnwidth]{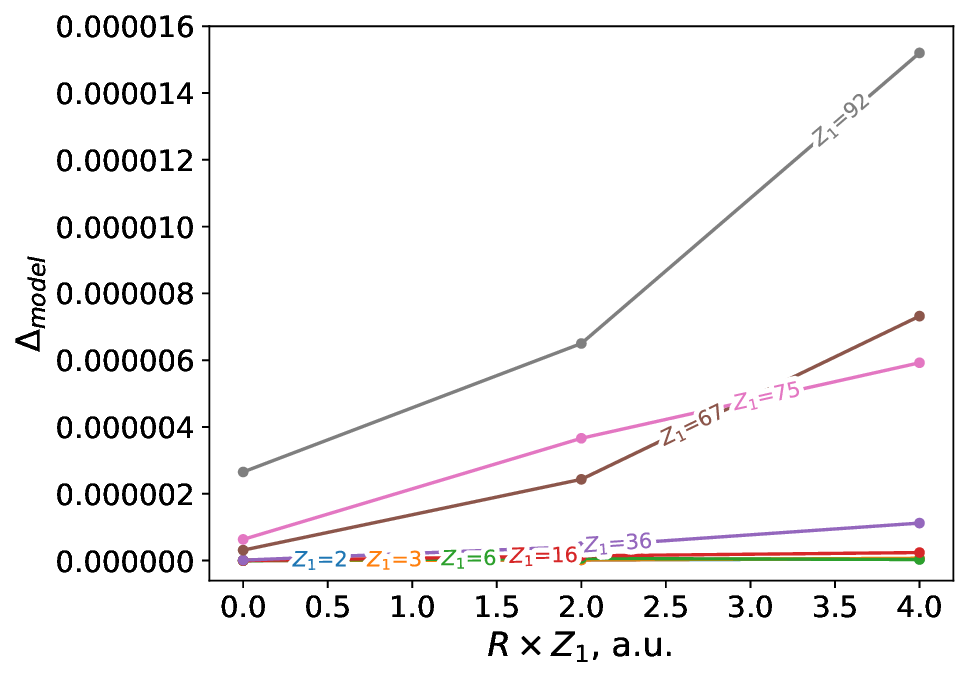}
\caption{The graphs showing the dependence of the relative magnitude $\Delta_{\mathrm{model}} =\left|E_{\mathrm{A-DKB}}^{(\rm{model}_1)}-E_{\mathrm{A-DKB}}^{(\rm{model}_2)}\right|/E_{\mathrm{A-DKB}}^{(\rm{model}_1)}$ at fixed nuclear radius and zero field on the inter-nuclear distance, $R$. }
\label{fig3}
\end{figure}

In turn, the dependence on the model of the nucleus was tested for the Fermi and spherical forms of the charge distribution of the nucleus in case of heavy ions and for spherical and shell models for light nuclei. Comparing the corresponding rows in different sectors of Table~\ref{tab:nm}, we find a discrepancy of numbers at the level of $10^{-5} - 10^{-6}$. Since the dependence on the nucleus model shows a larger scatter of values than the dependence on the charge radius, it can be concluded that collisions of antiprotons with hydrogen-like ions can be used to refine the parameters of the used models. The dependence on the inter-nuclear distance of the finite nuclear size model is shown in Fig.~\ref{fig3}.
In particular, the behaviour of the relative magnitude $\Delta_{\mathrm{model}} =\left|E_{\mathrm{A- DKB}}^{(\rm{model}_1)}-E_{\mathrm{A-DKB}}^{(\rm{model}_2)}\right|/E_{\mathrm{A-DKB}} ^{(\rm{model}_1)}$ clearly demonstrates this possibility. The final result for the binding energy should not depend on the utilised model.

Finally, it can be expected that the results obtained in the present study may be useful for the research areas discussed recently in \cite{Doser_2022}.

\section{Acknowledgement}
This work was supported by RSF grant \textnumero 23-22-00250.

\section{Data availability}             % такой раздел требуют в jcp

The data that support the findings of this study are available within the article. The data that support the findings of this study are available from the corresponding author upon reasonable request.

	\bibliography{mybibfile}

\begin{thebibliography}{62}%
\makeatletter
\providecommand \@ifxundefined [1]{%
 \@ifx{#1\undefined}
}%
\providecommand \@ifnum [1]{%
 \ifnum #1\expandafter \@firstoftwo
 \else \expandafter \@secondoftwo
 \fi
}%
\providecommand \@ifx [1]{%
 \ifx #1\expandafter \@firstoftwo
 \else \expandafter \@secondoftwo
 \fi
}%
\providecommand \natexlab [1]{#1}%
\providecommand \enquote  [1]{``#1''}%
\providecommand \bibnamefont  [1]{#1}%
\providecommand \bibfnamefont [1]{#1}%
\providecommand \citenamefont [1]{#1}%
\providecommand \href@noop [0]{\@secondoftwo}%
\providecommand \href [0]{\begingroup \@sanitize@url \@href}%
\providecommand \@href[1]{\@@startlink{#1}\@@href}%
\providecommand \@@href[1]{\endgroup#1\@@endlink}%
\providecommand \@sanitize@url [0]{\catcode `\\12\catcode `\$12\catcode `\&12\catcode `\#12\catcode `\^12\catcode `\_12\catcode `\%12\relax}%
\providecommand \@@startlink[1]{}%
\providecommand \@@endlink[0]{}%
\providecommand \url  [0]{\begingroup\@sanitize@url \@url }%
\providecommand \@url [1]{\endgroup\@href {#1}{\urlprefix }}%
\providecommand \urlprefix  [0]{URL }%
\providecommand \Eprint [0]{\href }%
\providecommand \doibase [0]{https://doi.org/}%
\providecommand \selectlanguage [0]{\@gobble}%
\providecommand \bibinfo  [0]{\@secondoftwo}%
\providecommand \bibfield  [0]{\@secondoftwo}%
\providecommand \translation [1]{[#1]}%
\providecommand \BibitemOpen [0]{}%
\providecommand \bibitemStop [0]{}%
\providecommand \bibitemNoStop [0]{.\EOS\space}%
\providecommand \EOS [0]{\spacefactor3000\relax}%
\providecommand \BibitemShut  [1]{\csname bibitem#1\endcsname}%
\let\auto@bib@innerbib\@empty
%</preamble>
\bibitem [{\citenamefont {Verd{\'u}}\ \emph {et~al.}(2004)\citenamefont {Verd{\'u}}, \citenamefont {Djeki{\'c}}, \citenamefont {Stahl}, \citenamefont {Valenzuela}, \citenamefont {Vogel}, \citenamefont {Werth}, \citenamefont {Kluge},\ and\ \citenamefont {Quint}}]{verdu2004determination}%
  \BibitemOpen
  \bibfield  {author} {\bibinfo {author} {\bibfnamefont {J.}~\bibnamefont {Verd{\'u}}}, \bibinfo {author} {\bibfnamefont {S.}~\bibnamefont {Djeki{\'c}}}, \bibinfo {author} {\bibfnamefont {S.}~\bibnamefont {Stahl}}, \bibinfo {author} {\bibfnamefont {T.}~\bibnamefont {Valenzuela}}, \bibinfo {author} {\bibfnamefont {M.}~\bibnamefont {Vogel}}, \bibinfo {author} {\bibfnamefont {G.}~\bibnamefont {Werth}}, \bibinfo {author} {\bibfnamefont {H.}~\bibnamefont {Kluge}},\ and\ \bibinfo {author} {\bibfnamefont {W.}~\bibnamefont {Quint}},\ }\bibfield  {title} {\bibinfo {title} {Determination of the g-factor of single hydrogen-like ions by mode coupling in a penning trap},\ }\href {https://doi.org/10.1238/Physica.Topical.112a00068} {\bibfield  {journal} {\bibinfo  {journal} {Physica Scripta}\ }\textbf {\bibinfo {volume} {2004}},\ \bibinfo {pages} {68} (\bibinfo {year} {2004})}\BibitemShut {NoStop}%
\bibitem [{\citenamefont {Beier}\ \emph {et~al.}(2003)\citenamefont {Beier}, \citenamefont {H{\"a}ffner}, \citenamefont {Hermanspahn}, \citenamefont {Djekic}, \citenamefont {Kluge}, \citenamefont {Quint}, \citenamefont {Stahl}, \citenamefont {Valenzuela}, \citenamefont {Verd{\'u}},\ and\ \citenamefont {Werth}}]{beier2003measurement}%
  \BibitemOpen
  \bibfield  {author} {\bibinfo {author} {\bibfnamefont {T.}~\bibnamefont {Beier}}, \bibinfo {author} {\bibfnamefont {H.}~\bibnamefont {H{\"a}ffner}}, \bibinfo {author} {\bibfnamefont {N.}~\bibnamefont {Hermanspahn}}, \bibinfo {author} {\bibfnamefont {S.}~\bibnamefont {Djekic}}, \bibinfo {author} {\bibfnamefont {H.-J.}\ \bibnamefont {Kluge}}, \bibinfo {author} {\bibfnamefont {W.}~\bibnamefont {Quint}}, \bibinfo {author} {\bibfnamefont {S.}~\bibnamefont {Stahl}}, \bibinfo {author} {\bibfnamefont {T.}~\bibnamefont {Valenzuela}}, \bibinfo {author} {\bibfnamefont {J.}~\bibnamefont {Verd{\'u}}},\ and\ \bibinfo {author} {\bibfnamefont {G.}~\bibnamefont {Werth}},\ }\bibfield  {title} {\bibinfo {title} {The measurement of the electronic g-factor in hydrogen-like ions—a promising tool for determining fundamental and nuclear constants},\ }in\ \href {https://doi.org/https://doi.org/10.1007/978-3-642-55560-2_7} {\emph {\bibinfo {booktitle} {Exotic Nuclei and Atomic Masses: Proceedings of the Third International
  Conference on Exotic Nuclei and Atomic Masses ENAM 2001 H{\"a}meenlinna, Finland, 2--7 July 2001}}}\ (\bibinfo {organization} {Springer},\ \bibinfo {year} {2003})\ pp.\ \bibinfo {pages} {29--32}\BibitemShut {NoStop}%
\bibitem [{\citenamefont {Vogel}\ \emph {et~al.}(2005)\citenamefont {Vogel}, \citenamefont {Alonso}, \citenamefont {Djekic}, \citenamefont {Kluge}, \citenamefont {Quint}, \citenamefont {Stahl}, \citenamefont {Verdu},\ and\ \citenamefont {Werth}}]{vogel2005towards}%
  \BibitemOpen
  \bibfield  {author} {\bibinfo {author} {\bibfnamefont {M.}~\bibnamefont {Vogel}}, \bibinfo {author} {\bibfnamefont {J.}~\bibnamefont {Alonso}}, \bibinfo {author} {\bibfnamefont {S.}~\bibnamefont {Djekic}}, \bibinfo {author} {\bibfnamefont {H.-J.}\ \bibnamefont {Kluge}}, \bibinfo {author} {\bibfnamefont {W.}~\bibnamefont {Quint}}, \bibinfo {author} {\bibfnamefont {S.}~\bibnamefont {Stahl}}, \bibinfo {author} {\bibfnamefont {J.}~\bibnamefont {Verdu}},\ and\ \bibinfo {author} {\bibfnamefont {G.}~\bibnamefont {Werth}},\ }\bibfield  {title} {\bibinfo {title} {Towards electronic g-factor measurements in medium-heavy hydrogen-like and lithium-like ions},\ }\href {https://doi.org/https://doi.org/10.1016/j.nimb.2005.03.136} {\bibfield  {journal} {\bibinfo  {journal} {Nuclear Instruments and Methods in Physics Research Section B: Beam Interactions with Materials and Atoms}\ }\textbf {\bibinfo {volume} {235}},\ \bibinfo {pages} {7} (\bibinfo {year} {2005})}\BibitemShut {NoStop}%
\bibitem [{\citenamefont {Diederich}\ \emph {et~al.}(1999)\citenamefont {Diederich}, \citenamefont {Häffner}, \citenamefont {Hermanspahn}, \citenamefont {Immel}, \citenamefont {Kluge}, \citenamefont {Ley}, \citenamefont {Mann}, \citenamefont {Stahl}, \citenamefont {Quint}, \citenamefont {Verdú},\ and\ \citenamefont {Werth}}]{10.1063/1.57483}%
  \BibitemOpen
  \bibfield  {author} {\bibinfo {author} {\bibfnamefont {M.}~\bibnamefont {Diederich}}, \bibinfo {author} {\bibfnamefont {H.}~\bibnamefont {Häffner}}, \bibinfo {author} {\bibfnamefont {N.}~\bibnamefont {Hermanspahn}}, \bibinfo {author} {\bibfnamefont {M.}~\bibnamefont {Immel}}, \bibinfo {author} {\bibfnamefont {H.~J.}\ \bibnamefont {Kluge}}, \bibinfo {author} {\bibfnamefont {R.}~\bibnamefont {Ley}}, \bibinfo {author} {\bibfnamefont {R.}~\bibnamefont {Mann}}, \bibinfo {author} {\bibfnamefont {S.}~\bibnamefont {Stahl}}, \bibinfo {author} {\bibfnamefont {W.}~\bibnamefont {Quint}}, \bibinfo {author} {\bibfnamefont {J.}~\bibnamefont {Verdú}},\ and\ \bibinfo {author} {\bibfnamefont {G.}~\bibnamefont {Werth}},\ }\bibfield  {title} {\bibinfo {title} {{The g-factor of hydrogen-like ions}},\ }\href {https://doi.org/10.1063/1.57483} {\bibfield  {journal} {\bibinfo  {journal} {AIP Conference Proceedings}\ }\textbf {\bibinfo {volume} {457}},\ \bibinfo {pages} {43} (\bibinfo {year} {1999})},\ \Eprint
  {https://arxiv.org/abs/https://pubs.aip.org/aip/acp/article-pdf/457/1/43/11807583/43\_1\_online.pdf} {https://pubs.aip.org/aip/acp/article-pdf/457/1/43/11807583/43\_1\_online.pdf} \BibitemShut {NoStop}%
\bibitem [{\citenamefont {Köhler}\ \emph {et~al.}(2015)\citenamefont {Köhler}, \citenamefont {Sturm}, \citenamefont {Kracke}, \citenamefont {Werth}, \citenamefont {Quint},\ and\ \citenamefont {Blaum}}]{Kohler_2015}%
  \BibitemOpen
  \bibfield  {author} {\bibinfo {author} {\bibfnamefont {F.}~\bibnamefont {Köhler}}, \bibinfo {author} {\bibfnamefont {S.}~\bibnamefont {Sturm}}, \bibinfo {author} {\bibfnamefont {A.}~\bibnamefont {Kracke}}, \bibinfo {author} {\bibfnamefont {G.}~\bibnamefont {Werth}}, \bibinfo {author} {\bibfnamefont {W.}~\bibnamefont {Quint}},\ and\ \bibinfo {author} {\bibfnamefont {K.}~\bibnamefont {Blaum}},\ }\bibfield  {title} {\bibinfo {title} {The electron mass from g-factor measurements on hydrogen-like carbon 12c5+},\ }\href {https://doi.org/10.1088/0953-4075/48/14/144032} {\bibfield  {journal} {\bibinfo  {journal} {Journal of Physics B: Atomic, Molecular and Optical Physics}\ }\textbf {\bibinfo {volume} {48}},\ \bibinfo {pages} {144032} (\bibinfo {year} {2015})}\BibitemShut {NoStop}%
\bibitem [{\citenamefont {Shabaev}\ \emph {et~al.}(2006)\citenamefont {Shabaev}, \citenamefont {Glazov}, \citenamefont {Oreshkina}, \citenamefont {Volotka}, \citenamefont {Plunien}, \citenamefont {Kluge},\ and\ \citenamefont {Quint}}]{PhysRevLett.96.253002}%
  \BibitemOpen
  \bibfield  {author} {\bibinfo {author} {\bibfnamefont {V.~M.}\ \bibnamefont {Shabaev}}, \bibinfo {author} {\bibfnamefont {D.~A.}\ \bibnamefont {Glazov}}, \bibinfo {author} {\bibfnamefont {N.~S.}\ \bibnamefont {Oreshkina}}, \bibinfo {author} {\bibfnamefont {A.~V.}\ \bibnamefont {Volotka}}, \bibinfo {author} {\bibfnamefont {G.}~\bibnamefont {Plunien}}, \bibinfo {author} {\bibfnamefont {H.-J.}\ \bibnamefont {Kluge}},\ and\ \bibinfo {author} {\bibfnamefont {W.}~\bibnamefont {Quint}},\ }\bibfield  {title} {\bibinfo {title} {$g$-factor of heavy ions: A new access to the fine structure constant},\ }\href {https://doi.org/10.1103/PhysRevLett.96.253002} {\bibfield  {journal} {\bibinfo  {journal} {Phys. Rev. Lett.}\ }\textbf {\bibinfo {volume} {96}},\ \bibinfo {pages} {253002} (\bibinfo {year} {2006})}\BibitemShut {NoStop}%
\bibitem [{\citenamefont {Yerokhin}\ \emph {et~al.}(2018)\citenamefont {Yerokhin}, \citenamefont {Pachucki},\ and\ \citenamefont {Patkóš}}]{https://doi.org/10.1002/andp.201800324}%
  \BibitemOpen
  \bibfield  {author} {\bibinfo {author} {\bibfnamefont {V.~A.}\ \bibnamefont {Yerokhin}}, \bibinfo {author} {\bibfnamefont {K.}~\bibnamefont {Pachucki}},\ and\ \bibinfo {author} {\bibfnamefont {V.}~\bibnamefont {Patkóš}},\ }\bibfield  {title} {\bibinfo {title} {Theory of the lamb shift in hydrogen and light hydrogen-like ions},\ }\href {https://doi.org/https://doi.org/10.1002/andp.201800324} {\bibfield  {journal} {\bibinfo  {journal} {Annalen der Physik}\ }\textbf {\bibinfo {volume} {531}},\ \bibinfo {pages} {1800324} (\bibinfo {year} {2018})},\ \Eprint {https://arxiv.org/abs/https://onlinelibrary.wiley.com/doi/pdf/10.1002/andp.201800324} {https://onlinelibrary.wiley.com/doi/pdf/10.1002/andp.201800324} \BibitemShut {NoStop}%
\bibitem [{\citenamefont {Andrianov}\ \emph {et~al.}(2009)\citenamefont {Andrianov}, \citenamefont {Beckert}, \citenamefont {Bleile}, \citenamefont {Chatterjee}, \citenamefont {Echler}, \citenamefont {Egelhof}, \citenamefont {Gumberidze}, \citenamefont {Ilieva}, \citenamefont {Kiselev}, \citenamefont {Kilbourne}, \citenamefont {Kluge}, \citenamefont {Kraft‐Bermuth}, \citenamefont {McCammon}, \citenamefont {Meier}, \citenamefont {Reuschl}, \citenamefont {Stöhlker},\ and\ \citenamefont {Trassinelli}}]{10.1063/1.3292459}%
  \BibitemOpen
  \bibfield  {author} {\bibinfo {author} {\bibfnamefont {V.}~\bibnamefont {Andrianov}}, \bibinfo {author} {\bibfnamefont {K.}~\bibnamefont {Beckert}}, \bibinfo {author} {\bibfnamefont {A.}~\bibnamefont {Bleile}}, \bibinfo {author} {\bibfnamefont {C.}~\bibnamefont {Chatterjee}}, \bibinfo {author} {\bibfnamefont {A.}~\bibnamefont {Echler}}, \bibinfo {author} {\bibfnamefont {P.}~\bibnamefont {Egelhof}}, \bibinfo {author} {\bibfnamefont {A.}~\bibnamefont {Gumberidze}}, \bibinfo {author} {\bibfnamefont {S.}~\bibnamefont {Ilieva}}, \bibinfo {author} {\bibfnamefont {O.}~\bibnamefont {Kiselev}}, \bibinfo {author} {\bibfnamefont {C.}~\bibnamefont {Kilbourne}}, \bibinfo {author} {\bibfnamefont {H.}~\bibnamefont {Kluge}}, \bibinfo {author} {\bibfnamefont {S.}~\bibnamefont {Kraft‐Bermuth}}, \bibinfo {author} {\bibfnamefont {D.}~\bibnamefont {McCammon}}, \bibinfo {author} {\bibfnamefont {J.~P.}\ \bibnamefont {Meier}}, \bibinfo {author} {\bibfnamefont {R.}~\bibnamefont {Reuschl}}, \bibinfo {author} {\bibfnamefont
  {T.}~\bibnamefont {Stöhlker}},\ and\ \bibinfo {author} {\bibfnamefont {M.}~\bibnamefont {Trassinelli}},\ }\bibfield  {title} {\bibinfo {title} {{Precise Lamb Shift Measurements in Hydrogen‐Like Heavy Ions—Status and Perspectives}},\ }\href {https://doi.org/https://doi.org/10.1063/1.3292459} {\bibfield  {journal} {\bibinfo  {journal} {AIP Conference Proceedings}\ }\textbf {\bibinfo {volume} {1185}},\ \bibinfo {pages} {99} (\bibinfo {year} {2009})}\BibitemShut {NoStop}%
\bibitem [{\citenamefont {Gassner}\ \emph {et~al.}(2018)\citenamefont {Gassner}, \citenamefont {Trassinelli}, \citenamefont {Heß}, \citenamefont {Spillmann}, \citenamefont {Banaś}, \citenamefont {Blumenhagen}, \citenamefont {Bosch}, \citenamefont {Brandau}, \citenamefont {Chen}, \citenamefont {Dimopoulou}, \citenamefont {Förster}, \citenamefont {Grisenti}, \citenamefont {Gumberidze}, \citenamefont {Hagmann}, \citenamefont {Hillenbrand}, \citenamefont {Indelicato}, \citenamefont {Jagodzinski}, \citenamefont {Kämpfer}, \citenamefont {Kozhuharov}, \citenamefont {Lestinsky}, \citenamefont {Liesen}, \citenamefont {Litvinov}, \citenamefont {Loetzsch}, \citenamefont {Manil}, \citenamefont {Märtin}, \citenamefont {Nolden}, \citenamefont {Petridis}, \citenamefont {Sanjari}, \citenamefont {Schulze}, \citenamefont {Schwemlein}, \citenamefont {Simionovici}, \citenamefont {Steck}, \citenamefont {Stöhlker}, \citenamefont {Szabo}, \citenamefont {Trotsenko}, \citenamefont {Uschmann}, \citenamefont {Weber},
  \citenamefont {Wehrhan}, \citenamefont {Winckler}, \citenamefont {Winters}, \citenamefont {Winters}, \citenamefont {Ziegler},\ and\ \citenamefont {Beyer}}]{Gassner_2018}%
  \BibitemOpen
  \bibfield  {author} {\bibinfo {author} {\bibfnamefont {T.}~\bibnamefont {Gassner}}, \bibinfo {author} {\bibfnamefont {M.}~\bibnamefont {Trassinelli}}, \bibinfo {author} {\bibfnamefont {R.}~\bibnamefont {Heß}}, \bibinfo {author} {\bibfnamefont {U.}~\bibnamefont {Spillmann}}, \bibinfo {author} {\bibfnamefont {D.}~\bibnamefont {Banaś}}, \bibinfo {author} {\bibfnamefont {K.-H.}\ \bibnamefont {Blumenhagen}}, \bibinfo {author} {\bibfnamefont {F.}~\bibnamefont {Bosch}}, \bibinfo {author} {\bibfnamefont {C.}~\bibnamefont {Brandau}}, \bibinfo {author} {\bibfnamefont {W.}~\bibnamefont {Chen}}, \bibinfo {author} {\bibfnamefont {C.}~\bibnamefont {Dimopoulou}}, \bibinfo {author} {\bibfnamefont {E.}~\bibnamefont {Förster}}, \bibinfo {author} {\bibfnamefont {R.~E.}\ \bibnamefont {Grisenti}}, \bibinfo {author} {\bibfnamefont {A.}~\bibnamefont {Gumberidze}}, \bibinfo {author} {\bibfnamefont {S.}~\bibnamefont {Hagmann}}, \bibinfo {author} {\bibfnamefont {P.-M.}\ \bibnamefont {Hillenbrand}}, \bibinfo {author}
  {\bibfnamefont {P.}~\bibnamefont {Indelicato}}, \bibinfo {author} {\bibfnamefont {P.}~\bibnamefont {Jagodzinski}}, \bibinfo {author} {\bibfnamefont {T.}~\bibnamefont {Kämpfer}}, \bibinfo {author} {\bibfnamefont {C.}~\bibnamefont {Kozhuharov}}, \bibinfo {author} {\bibfnamefont {M.}~\bibnamefont {Lestinsky}}, \bibinfo {author} {\bibfnamefont {D.}~\bibnamefont {Liesen}}, \bibinfo {author} {\bibfnamefont {Y.~A.}\ \bibnamefont {Litvinov}}, \bibinfo {author} {\bibfnamefont {R.}~\bibnamefont {Loetzsch}}, \bibinfo {author} {\bibfnamefont {B.}~\bibnamefont {Manil}}, \bibinfo {author} {\bibfnamefont {R.}~\bibnamefont {Märtin}}, \bibinfo {author} {\bibfnamefont {F.}~\bibnamefont {Nolden}}, \bibinfo {author} {\bibfnamefont {N.}~\bibnamefont {Petridis}}, \bibinfo {author} {\bibfnamefont {M.~S.}\ \bibnamefont {Sanjari}}, \bibinfo {author} {\bibfnamefont {K.~S.}\ \bibnamefont {Schulze}}, \bibinfo {author} {\bibfnamefont {M.}~\bibnamefont {Schwemlein}}, \bibinfo {author} {\bibfnamefont {A.}~\bibnamefont {Simionovici}},
  \bibinfo {author} {\bibfnamefont {M.}~\bibnamefont {Steck}}, \bibinfo {author} {\bibfnamefont {T.}~\bibnamefont {Stöhlker}}, \bibinfo {author} {\bibfnamefont {C.~I.}\ \bibnamefont {Szabo}}, \bibinfo {author} {\bibfnamefont {S.}~\bibnamefont {Trotsenko}}, \bibinfo {author} {\bibfnamefont {I.}~\bibnamefont {Uschmann}}, \bibinfo {author} {\bibfnamefont {G.}~\bibnamefont {Weber}}, \bibinfo {author} {\bibfnamefont {O.}~\bibnamefont {Wehrhan}}, \bibinfo {author} {\bibfnamefont {N.}~\bibnamefont {Winckler}}, \bibinfo {author} {\bibfnamefont {D.~F.~A.}\ \bibnamefont {Winters}}, \bibinfo {author} {\bibfnamefont {N.}~\bibnamefont {Winters}}, \bibinfo {author} {\bibfnamefont {E.}~\bibnamefont {Ziegler}},\ and\ \bibinfo {author} {\bibfnamefont {H.~F.}\ \bibnamefont {Beyer}},\ }\bibfield  {title} {\bibinfo {title} {Wavelength-dispersive spectroscopy in the hard x-ray regime of a heavy highly-charged ion: the 1s lamb shift in hydrogen-like gold},\ }\href {https://doi.org/10.1088/1367-2630/aad01d} {\bibfield  {journal}
  {\bibinfo  {journal} {New Journal of Physics}\ }\textbf {\bibinfo {volume} {20}},\ \bibinfo {pages} {073033} (\bibinfo {year} {2018})}\BibitemShut {NoStop}%
\bibitem [{\citenamefont {Kraft-Bermuth}\ \emph {et~al.}(2017)\citenamefont {Kraft-Bermuth}, \citenamefont {Andrianov}, \citenamefont {Bleile}, \citenamefont {Echler}, \citenamefont {Egelhof}, \citenamefont {Grabitz}, \citenamefont {Ilieva}, \citenamefont {Kiselev}, \citenamefont {Kilbourne}, \citenamefont {McCammon}, \citenamefont {Meier},\ and\ \citenamefont {Scholz}}]{Kraft-Bermuth_2017}%
  \BibitemOpen
  \bibfield  {author} {\bibinfo {author} {\bibfnamefont {S.}~\bibnamefont {Kraft-Bermuth}}, \bibinfo {author} {\bibfnamefont {V.}~\bibnamefont {Andrianov}}, \bibinfo {author} {\bibfnamefont {A.}~\bibnamefont {Bleile}}, \bibinfo {author} {\bibfnamefont {A.}~\bibnamefont {Echler}}, \bibinfo {author} {\bibfnamefont {P.}~\bibnamefont {Egelhof}}, \bibinfo {author} {\bibfnamefont {P.}~\bibnamefont {Grabitz}}, \bibinfo {author} {\bibfnamefont {S.}~\bibnamefont {Ilieva}}, \bibinfo {author} {\bibfnamefont {O.}~\bibnamefont {Kiselev}}, \bibinfo {author} {\bibfnamefont {C.}~\bibnamefont {Kilbourne}}, \bibinfo {author} {\bibfnamefont {D.}~\bibnamefont {McCammon}}, \bibinfo {author} {\bibfnamefont {J.~P.}\ \bibnamefont {Meier}},\ and\ \bibinfo {author} {\bibfnamefont {P.}~\bibnamefont {Scholz}},\ }\bibfield  {title} {\bibinfo {title} {Precise determination of the 1s lamb shift in hydrogen-like lead and gold using microcalorimeters},\ }\href {https://doi.org/10.1088/1361-6455/50/5/055603} {\bibfield  {journal} {\bibinfo
  {journal} {Journal of Physics B: Atomic, Molecular and Optical Physics}\ }\textbf {\bibinfo {volume} {50}},\ \bibinfo {pages} {055603} (\bibinfo {year} {2017})}\BibitemShut {NoStop}%
\bibitem [{\citenamefont {Crespo L\'opez-Urrutia}\ \emph {et~al.}(1998)\citenamefont {Crespo L\'opez-Urrutia}, \citenamefont {Beiersdorfer}, \citenamefont {Widmann}, \citenamefont {Birkett}, \citenamefont {M\aa{}rtensson-Pendrill},\ and\ \citenamefont {Gustavsson}}]{PhysRevA.57.879}%
  \BibitemOpen
  \bibfield  {author} {\bibinfo {author} {\bibfnamefont {J.~R.}\ \bibnamefont {Crespo L\'opez-Urrutia}}, \bibinfo {author} {\bibfnamefont {P.}~\bibnamefont {Beiersdorfer}}, \bibinfo {author} {\bibfnamefont {K.}~\bibnamefont {Widmann}}, \bibinfo {author} {\bibfnamefont {B.~B.}\ \bibnamefont {Birkett}}, \bibinfo {author} {\bibfnamefont {A.-M.}\ \bibnamefont {M\aa{}rtensson-Pendrill}},\ and\ \bibinfo {author} {\bibfnamefont {M.~G.~H.}\ \bibnamefont {Gustavsson}},\ }\bibfield  {title} {\bibinfo {title} {Nuclear magnetization distribution radii determined by hyperfine transitions in the $1s$ level of h-like ions ${}^{185}{\mathrm{re}}^{74+}$ and ${}^{187}{\mathrm{re}}^{74+}$},\ }\href {https://doi.org/10.1103/PhysRevA.57.879} {\bibfield  {journal} {\bibinfo  {journal} {Phys. Rev. A}\ }\textbf {\bibinfo {volume} {57}},\ \bibinfo {pages} {879} (\bibinfo {year} {1998})}\BibitemShut {NoStop}%
\bibitem [{\citenamefont {Klaft}\ \emph {et~al.}(1994)\citenamefont {Klaft}, \citenamefont {Borneis}, \citenamefont {Engel}, \citenamefont {Fricke}, \citenamefont {Grieser}, \citenamefont {Huber}, \citenamefont {K\"uhl}, \citenamefont {Marx}, \citenamefont {Neumann}, \citenamefont {Schr\"oder}, \citenamefont {Seelig},\ and\ \citenamefont {V\"olker}}]{PhysRevLett.73.2425}%
  \BibitemOpen
  \bibfield  {author} {\bibinfo {author} {\bibfnamefont {I.}~\bibnamefont {Klaft}}, \bibinfo {author} {\bibfnamefont {S.}~\bibnamefont {Borneis}}, \bibinfo {author} {\bibfnamefont {T.}~\bibnamefont {Engel}}, \bibinfo {author} {\bibfnamefont {B.}~\bibnamefont {Fricke}}, \bibinfo {author} {\bibfnamefont {R.}~\bibnamefont {Grieser}}, \bibinfo {author} {\bibfnamefont {G.}~\bibnamefont {Huber}}, \bibinfo {author} {\bibfnamefont {T.}~\bibnamefont {K\"uhl}}, \bibinfo {author} {\bibfnamefont {D.}~\bibnamefont {Marx}}, \bibinfo {author} {\bibfnamefont {R.}~\bibnamefont {Neumann}}, \bibinfo {author} {\bibfnamefont {S.}~\bibnamefont {Schr\"oder}}, \bibinfo {author} {\bibfnamefont {P.}~\bibnamefont {Seelig}},\ and\ \bibinfo {author} {\bibfnamefont {L.}~\bibnamefont {V\"olker}},\ }\bibfield  {title} {\bibinfo {title} {Precision laser spectroscopy of the ground state hyperfine splitting of hydrogenlike $^{209}\mathrm{Bi}^{82+}$},\ }\href {https://doi.org/10.1103/PhysRevLett.73.2425} {\bibfield  {journal} {\bibinfo
  {journal} {Phys. Rev. Lett.}\ }\textbf {\bibinfo {volume} {73}},\ \bibinfo {pages} {2425} (\bibinfo {year} {1994})}\BibitemShut {NoStop}%
\bibitem [{\citenamefont {Gershtein}\ and\ \citenamefont {Zeldovich}(1970)}]{gershtein1970positron}%
  \BibitemOpen
  \bibfield  {author} {\bibinfo {author} {\bibfnamefont {S.}~\bibnamefont {Gershtein}}\ and\ \bibinfo {author} {\bibfnamefont {Y.}~\bibnamefont {Zeldovich}},\ }\bibfield  {title} {\bibinfo {title} {Positron production during the mutual approach of heavy nuclei and the polarization of the vacuum},\ }\href {http://jetp.ras.ru/cgi-bin/dn/e_030_02_0358.pdf} {\bibfield  {journal} {\bibinfo  {journal} {Sov. Phys. JETP}\ }\textbf {\bibinfo {volume} {30}},\ \bibinfo {pages} {358} (\bibinfo {year} {1970})}\BibitemShut {NoStop}%
\bibitem [{\citenamefont {Pieper}\ and\ \citenamefont {Greiner}(1969)}]{pieper1969interior}%
  \BibitemOpen
  \bibfield  {author} {\bibinfo {author} {\bibfnamefont {W.}~\bibnamefont {Pieper}}\ and\ \bibinfo {author} {\bibfnamefont {W.}~\bibnamefont {Greiner}},\ }\bibfield  {title} {\bibinfo {title} {Interior electron shells in superheavy nuclei},\ }\href {https://doi.org/https://doi.org/10.1007/BF01670014} {\bibfield  {journal} {\bibinfo  {journal} {Zeitschrift f{\"u}r Physik A Hadrons and nuclei}\ }\textbf {\bibinfo {volume} {218}},\ \bibinfo {pages} {327} (\bibinfo {year} {1969})}\BibitemShut {NoStop}%
\bibitem [{\citenamefont {Zeldovich}\ and\ \citenamefont {Popov}(1972)}]{Ya_B_Zeldovich_1972}%
  \BibitemOpen
  \bibfield  {author} {\bibinfo {author} {\bibfnamefont {Y.~B.}\ \bibnamefont {Zeldovich}}\ and\ \bibinfo {author} {\bibfnamefont {V.~S.}\ \bibnamefont {Popov}},\ }\bibfield  {title} {\bibinfo {title} {Electronic structure of superheavy atoms},\ }\href {https://doi.org/10.1070/PU1972v014n06ABEH004735} {\bibfield  {journal} {\bibinfo  {journal} {Soviet Physics Uspekhi}\ }\textbf {\bibinfo {volume} {14}},\ \bibinfo {pages} {673} (\bibinfo {year} {1972})}\BibitemShut {NoStop}%
\bibitem [{\citenamefont {Rafelski}\ \emph {et~al.}(1978)\citenamefont {Rafelski}, \citenamefont {Fulcher},\ and\ \citenamefont {Klein}}]{RAFELSKI1978227}%
  \BibitemOpen
  \bibfield  {author} {\bibinfo {author} {\bibfnamefont {J.}~\bibnamefont {Rafelski}}, \bibinfo {author} {\bibfnamefont {L.~P.}\ \bibnamefont {Fulcher}},\ and\ \bibinfo {author} {\bibfnamefont {A.}~\bibnamefont {Klein}},\ }\bibfield  {title} {\bibinfo {title} {Fermions and bosons interacting with arbitrarily strong external fields},\ }\href {https://doi.org/https://doi.org/10.1016/0370-1573(78)90116-3} {\bibfield  {journal} {\bibinfo  {journal} {Physics Reports}\ }\textbf {\bibinfo {volume} {38}},\ \bibinfo {pages} {227} (\bibinfo {year} {1978})}\BibitemShut {NoStop}%
\bibitem [{\citenamefont {Shabaev}\ \emph {et~al.}(2019)\citenamefont {Shabaev}, \citenamefont {Bondarev}, \citenamefont {Glazov}, \citenamefont {Kozhedub}, \citenamefont {Maltsev}, \citenamefont {Malyshev}, \citenamefont {Popov}, \citenamefont {Tumakov},\ and\ \citenamefont {Tupitsyn}}]{shabaev2019qed}%
  \BibitemOpen
  \bibfield  {author} {\bibinfo {author} {\bibfnamefont {V.}~\bibnamefont {Shabaev}}, \bibinfo {author} {\bibfnamefont {A.}~\bibnamefont {Bondarev}}, \bibinfo {author} {\bibfnamefont {D.}~\bibnamefont {Glazov}}, \bibinfo {author} {\bibfnamefont {Y.}~\bibnamefont {Kozhedub}}, \bibinfo {author} {\bibfnamefont {I.}~\bibnamefont {Maltsev}}, \bibinfo {author} {\bibfnamefont {A.}~\bibnamefont {Malyshev}}, \bibinfo {author} {\bibfnamefont {R.}~\bibnamefont {Popov}}, \bibinfo {author} {\bibfnamefont {D.}~\bibnamefont {Tumakov}},\ and\ \bibinfo {author} {\bibfnamefont {I.}~\bibnamefont {Tupitsyn}},\ }\bibfield  {title} {\bibinfo {title} {Qed with heavy ions: on the way from strong to supercritical fields},\ }\href {https://arxiv.org/abs/1910.01373} {\bibfield  {journal} {\bibinfo  {journal} {arXiv preprint arXiv:1910.01373}\ } (\bibinfo {year} {2019})}\BibitemShut {NoStop}%
\bibitem [{\citenamefont {Maltsev}\ \emph {et~al.}(2019)\citenamefont {Maltsev}, \citenamefont {Shabaev}, \citenamefont {Popov}, \citenamefont {Kozhedub}, \citenamefont {Plunien}, \citenamefont {Ma}, \citenamefont {St\"ohlker},\ and\ \citenamefont {Tumakov}}]{PhysRevLett.123.113401}%
  \BibitemOpen
  \bibfield  {author} {\bibinfo {author} {\bibfnamefont {I.~A.}\ \bibnamefont {Maltsev}}, \bibinfo {author} {\bibfnamefont {V.~M.}\ \bibnamefont {Shabaev}}, \bibinfo {author} {\bibfnamefont {R.~V.}\ \bibnamefont {Popov}}, \bibinfo {author} {\bibfnamefont {Y.~S.}\ \bibnamefont {Kozhedub}}, \bibinfo {author} {\bibfnamefont {G.}~\bibnamefont {Plunien}}, \bibinfo {author} {\bibfnamefont {X.}~\bibnamefont {Ma}}, \bibinfo {author} {\bibfnamefont {T.}~\bibnamefont {St\"ohlker}},\ and\ \bibinfo {author} {\bibfnamefont {D.~A.}\ \bibnamefont {Tumakov}},\ }\bibfield  {title} {\bibinfo {title} {How to observe the vacuum decay in low-energy heavy-ion collisions},\ }\href {https://doi.org/10.1103/PhysRevLett.123.113401} {\bibfield  {journal} {\bibinfo  {journal} {Phys. Rev. Lett.}\ }\textbf {\bibinfo {volume} {123}},\ \bibinfo {pages} {113401} (\bibinfo {year} {2019})}\BibitemShut {NoStop}%
\bibitem [{\citenamefont {Popov}\ \emph {et~al.}(2020)\citenamefont {Popov}, \citenamefont {Shabaev}, \citenamefont {Telnov}, \citenamefont {Tupitsyn}, \citenamefont {Maltsev}, \citenamefont {Kozhedub}, \citenamefont {Bondarev}, \citenamefont {Kozin}, \citenamefont {Ma}, \citenamefont {Plunien}, \citenamefont {St\"ohlker}, \citenamefont {Tumakov},\ and\ \citenamefont {Zaytsev}}]{PhysRevD.102.076005}%
  \BibitemOpen
  \bibfield  {author} {\bibinfo {author} {\bibfnamefont {R.~V.}\ \bibnamefont {Popov}}, \bibinfo {author} {\bibfnamefont {V.~M.}\ \bibnamefont {Shabaev}}, \bibinfo {author} {\bibfnamefont {D.~A.}\ \bibnamefont {Telnov}}, \bibinfo {author} {\bibfnamefont {I.~I.}\ \bibnamefont {Tupitsyn}}, \bibinfo {author} {\bibfnamefont {I.~A.}\ \bibnamefont {Maltsev}}, \bibinfo {author} {\bibfnamefont {Y.~S.}\ \bibnamefont {Kozhedub}}, \bibinfo {author} {\bibfnamefont {A.~I.}\ \bibnamefont {Bondarev}}, \bibinfo {author} {\bibfnamefont {N.~V.}\ \bibnamefont {Kozin}}, \bibinfo {author} {\bibfnamefont {X.}~\bibnamefont {Ma}}, \bibinfo {author} {\bibfnamefont {G.}~\bibnamefont {Plunien}}, \bibinfo {author} {\bibfnamefont {T.}~\bibnamefont {St\"ohlker}}, \bibinfo {author} {\bibfnamefont {D.~A.}\ \bibnamefont {Tumakov}},\ and\ \bibinfo {author} {\bibfnamefont {V.~A.}\ \bibnamefont {Zaytsev}},\ }\bibfield  {title} {\bibinfo {title} {How to access qed at a supercritical coulomb field},\ }\href
  {https://doi.org/10.1103/PhysRevD.102.076005} {\bibfield  {journal} {\bibinfo  {journal} {Phys. Rev. D}\ }\textbf {\bibinfo {volume} {102}},\ \bibinfo {pages} {076005} (\bibinfo {year} {2020})}\BibitemShut {NoStop}%
\bibitem [{\citenamefont {Voskresensky}(2021)}]{universe7040104}%
  \BibitemOpen
  \bibfield  {author} {\bibinfo {author} {\bibfnamefont {D.~N.}\ \bibnamefont {Voskresensky}},\ }\bibfield  {title} {\bibinfo {title} {Electron-positron vacuum instability in strong electric fields. relativistic semiclassical approach},\ }\bibfield  {journal} {\bibinfo  {journal} {Universe}\ }\textbf {\bibinfo {volume} {7}},\ \href {https://doi.org/10.3390/universe7040104} {10.3390/universe7040104} (\bibinfo {year} {2021})\BibitemShut {NoStop}%
\bibitem [{\citenamefont {Kotov}\ \emph {et~al.}(2020)\citenamefont {Kotov}, \citenamefont {Glazov}, \citenamefont {Malyshev}, \citenamefont {Vladimirova}, \citenamefont {Shabaev},\ and\ \citenamefont {Plunien}}]{kotov2020ground}%
  \BibitemOpen
  \bibfield  {author} {\bibinfo {author} {\bibfnamefont {A.~A.}\ \bibnamefont {Kotov}}, \bibinfo {author} {\bibfnamefont {D.~A.}\ \bibnamefont {Glazov}}, \bibinfo {author} {\bibfnamefont {A.~V.}\ \bibnamefont {Malyshev}}, \bibinfo {author} {\bibfnamefont {A.~V.}\ \bibnamefont {Vladimirova}}, \bibinfo {author} {\bibfnamefont {V.~M.}\ \bibnamefont {Shabaev}},\ and\ \bibinfo {author} {\bibfnamefont {G.}~\bibnamefont {Plunien}},\ }\bibfield  {title} {\bibinfo {title} {Ground-state energy of uranium diatomic quasimolecules with one and two electrons},\ }\href {https://doi.org/https://doi.org/10.1002/xrs.3064} {\bibfield  {journal} {\bibinfo  {journal} {X-Ray Spectrometry}\ }\textbf {\bibinfo {volume} {49}},\ \bibinfo {pages} {110} (\bibinfo {year} {2020})}\BibitemShut {NoStop}%
\bibitem [{\citenamefont {Kotov}\ \emph {et~al.}(2021)\citenamefont {Kotov}, \citenamefont {Glazov}, \citenamefont {Shabaev},\ and\ \citenamefont {Plunien}}]{kotov2021one}%
  \BibitemOpen
  \bibfield  {author} {\bibinfo {author} {\bibfnamefont {A.~A.}\ \bibnamefont {Kotov}}, \bibinfo {author} {\bibfnamefont {D.~A.}\ \bibnamefont {Glazov}}, \bibinfo {author} {\bibfnamefont {V.~M.}\ \bibnamefont {Shabaev}},\ and\ \bibinfo {author} {\bibfnamefont {G.}~\bibnamefont {Plunien}},\ }\bibfield  {title} {\bibinfo {title} {One-electron energy spectra of heavy highly charged quasimolecules: Finite-basis-set approach},\ }\href {https://doi.org/https://doi.org/10.3390/atoms9030044} {\bibfield  {journal} {\bibinfo  {journal} {Atoms}\ }\textbf {\bibinfo {volume} {9}},\ \bibinfo {pages} {44} (\bibinfo {year} {2021})}\BibitemShut {NoStop}%
\bibitem [{\citenamefont {Kotov}\ \emph {et~al.}(2022)\citenamefont {Kotov}, \citenamefont {Glazov}, \citenamefont {Malyshev}, \citenamefont {Shabaev},\ and\ \citenamefont {Plunien}}]{kotov2022}%
  \BibitemOpen
  \bibfield  {author} {\bibinfo {author} {\bibfnamefont {A.~A.}\ \bibnamefont {Kotov}}, \bibinfo {author} {\bibfnamefont {D.~A.}\ \bibnamefont {Glazov}}, \bibinfo {author} {\bibfnamefont {A.~V.}\ \bibnamefont {Malyshev}}, \bibinfo {author} {\bibfnamefont {V.~M.}\ \bibnamefont {Shabaev}},\ and\ \bibinfo {author} {\bibfnamefont {G.}~\bibnamefont {Plunien}},\ }\bibfield  {title} {\bibinfo {title} {Finite-basis-set approach to the two-center heteronuclear dirac problem},\ }\href {https://doi.org/10.3390/atoms10040145} {\bibfield  {journal} {\bibinfo  {journal} {Atoms}\ }\textbf {\bibinfo {volume} {10}},\ \bibinfo {pages} {145} (\bibinfo {year} {2022})}\BibitemShut {NoStop}%
\bibitem [{\citenamefont {Jonsell}(2018)}]{doi:10.1098/rsta.2017.0271}%
  \BibitemOpen
  \bibfield  {author} {\bibinfo {author} {\bibfnamefont {S.}~\bibnamefont {Jonsell}},\ }\bibfield  {title} {\bibinfo {title} {Collisions involving antiprotons and antihydrogen: an overview},\ }\href {https://doi.org/10.1098/rsta.2017.0271} {\bibfield  {journal} {\bibinfo  {journal} {Philosophical Transactions of the Royal Society A: Mathematical, Physical and Engineering Sciences}\ }\textbf {\bibinfo {volume} {376}},\ \bibinfo {pages} {20170271} (\bibinfo {year} {2018})}\BibitemShut {NoStop}%
\bibitem [{\citenamefont {Kirchner}\ and\ \citenamefont {Knudsen}(2011)}]{kirchner2011current}%
  \BibitemOpen
  \bibfield  {author} {\bibinfo {author} {\bibfnamefont {T.}~\bibnamefont {Kirchner}}\ and\ \bibinfo {author} {\bibfnamefont {H.}~\bibnamefont {Knudsen}},\ }\bibfield  {title} {\bibinfo {title} {Current status of antiproton impact ionization of atoms and molecules: theoretical and experimental perspectives},\ }\href {https://doi.org/10.1088/0953-4075/44/12/122001} {\bibfield  {journal} {\bibinfo  {journal} {Journal of Physics B: Atomic, Molecular and Optical Physics}\ }\textbf {\bibinfo {volume} {44}},\ \bibinfo {pages} {122001} (\bibinfo {year} {2011})}\BibitemShut {NoStop}%
\bibitem [{\citenamefont {Sakimoto}(2004)}]{sakimoto2004protonium}%
  \BibitemOpen
  \bibfield  {author} {\bibinfo {author} {\bibfnamefont {K.}~\bibnamefont {Sakimoto}},\ }\bibfield  {title} {\bibinfo {title} {Protonium formation in collisions of antiprotons with hydrogen molecular ions},\ }\href {https://doi.org/10.1088/0953-4075/37/11/004} {\bibfield  {journal} {\bibinfo  {journal} {Journal of Physics B: Atomic, Molecular and Optical Physics}\ }\textbf {\bibinfo {volume} {37}},\ \bibinfo {pages} {2255} (\bibinfo {year} {2004})}\BibitemShut {NoStop}%
\bibitem [{\citenamefont {Myers}(2018)}]{PhysRevA.98.010101}%
  \BibitemOpen
  \bibfield  {author} {\bibinfo {author} {\bibfnamefont {E.~G.}\ \bibnamefont {Myers}},\ }\bibfield  {title} {\bibinfo {title} {$cpt$ tests with the antihydrogen molecular ion},\ }\href {https://doi.org/10.1103/PhysRevA.98.010101} {\bibfield  {journal} {\bibinfo  {journal} {Phys. Rev. A}\ }\textbf {\bibinfo {volume} {98}},\ \bibinfo {pages} {010101} (\bibinfo {year} {2018})}\BibitemShut {NoStop}%
\bibitem [{\citenamefont {Genkin}\ and\ \citenamefont {Lindroth}(2009)}]{genkin2009possibility}%
  \BibitemOpen
  \bibfield  {author} {\bibinfo {author} {\bibfnamefont {M.}~\bibnamefont {Genkin}}\ and\ \bibinfo {author} {\bibfnamefont {E.}~\bibnamefont {Lindroth}},\ }\bibfield  {title} {\bibinfo {title} {Possibility of resonant capture of antiprotons by highly charged hydrogenlike ions},\ }\href {https://doi.org/https://doi.org/10.1140/epjd/e2008-00273-1} {\bibfield  {journal} {\bibinfo  {journal} {The European Physical Journal D}\ }\textbf {\bibinfo {volume} {51}},\ \bibinfo {pages} {205} (\bibinfo {year} {2009})}\BibitemShut {NoStop}%
\bibitem [{\citenamefont {Demkov}\ \emph {et~al.}(1984)\citenamefont {Demkov}, \citenamefont {Ostrovskii},\ and\ \citenamefont {Tel’nov}}]{demkov1984new}%
  \BibitemOpen
  \bibfield  {author} {\bibinfo {author} {\bibfnamefont {Y.~N.}\ \bibnamefont {Demkov}}, \bibinfo {author} {\bibfnamefont {V.}~\bibnamefont {Ostrovskii}},\ and\ \bibinfo {author} {\bibfnamefont {D.}~\bibnamefont {Tel’nov}},\ }\bibfield  {title} {\bibinfo {title} {New type of cross section singularity in backward scattering: the coulomb glory},\ }\href {http://www.jetp.ras.ru/cgi-bin/dn/e_059_02_0257.pdf} {\bibfield  {journal} {\bibinfo  {journal} {Zh. Eksp. Teor. Fiz.}\ }\textbf {\bibinfo {volume} {86}},\ \bibinfo {pages} {450} (\bibinfo {year} {1984})}\BibitemShut {NoStop}%
\bibitem [{\citenamefont {Demkov}\ and\ \citenamefont {Ostrovsky}(2001)}]{Yu_N_Demkov_2001}%
  \BibitemOpen
  \bibfield  {author} {\bibinfo {author} {\bibfnamefont {Y.~N.}\ \bibnamefont {Demkov}}\ and\ \bibinfo {author} {\bibfnamefont {V.~N.}\ \bibnamefont {Ostrovsky}},\ }\bibfield  {title} {\bibinfo {title} {Enhanced backscattering in antiproton-atom collision: Coulomb glory},\ }\href {https://doi.org/10.1088/0953-4075/34/18/102} {\bibfield  {journal} {\bibinfo  {journal} {Journal of Physics B: Atomic, Molecular and Optical Physics}\ }\textbf {\bibinfo {volume} {34}},\ \bibinfo {pages} {L595} (\bibinfo {year} {2001})}\BibitemShut {NoStop}%
\bibitem [{\citenamefont {Maiorova}\ \emph {et~al.}(2007)\citenamefont {Maiorova}, \citenamefont {Telnov}, \citenamefont {Shabaev}, \citenamefont {Tupitsyn}, \citenamefont {Plunien},\ and\ \citenamefont {St\"ohlker}}]{PhysRevA.76.032709}%
  \BibitemOpen
  \bibfield  {author} {\bibinfo {author} {\bibfnamefont {A.~V.}\ \bibnamefont {Maiorova}}, \bibinfo {author} {\bibfnamefont {D.~A.}\ \bibnamefont {Telnov}}, \bibinfo {author} {\bibfnamefont {V.~M.}\ \bibnamefont {Shabaev}}, \bibinfo {author} {\bibfnamefont {I.~I.}\ \bibnamefont {Tupitsyn}}, \bibinfo {author} {\bibfnamefont {G.}~\bibnamefont {Plunien}},\ and\ \bibinfo {author} {\bibfnamefont {T.}~\bibnamefont {St\"ohlker}},\ }\bibfield  {title} {\bibinfo {title} {Backward scattering of low-energy antiprotons by highly charged and neutral uranium: Coulomb glory},\ }\href {https://doi.org/10.1103/PhysRevA.76.032709} {\bibfield  {journal} {\bibinfo  {journal} {Phys. Rev. A}\ }\textbf {\bibinfo {volume} {76}},\ \bibinfo {pages} {032709} (\bibinfo {year} {2007})}\BibitemShut {NoStop}%
\bibitem [{\citenamefont {Korobov}\ and\ \citenamefont {Tsogbayar}(2007)}]{korobov2007relativistic1}%
  \BibitemOpen
  \bibfield  {author} {\bibinfo {author} {\bibfnamefont {V.}~\bibnamefont {Korobov}}\ and\ \bibinfo {author} {\bibfnamefont {T.}~\bibnamefont {Tsogbayar}},\ }\bibfield  {title} {\bibinfo {title} {Relativistic corrections of order m$\alpha$6 to the two-centre problem},\ }\href {https://doi.org/10.1088/0953-4075/40/13/011} {\bibfield  {journal} {\bibinfo  {journal} {Journal of Physics B: Atomic, Molecular and Optical Physics}\ }\textbf {\bibinfo {volume} {40}},\ \bibinfo {pages} {2661} (\bibinfo {year} {2007})}\BibitemShut {NoStop}%
\bibitem [{\citenamefont {Tsogbayar}\ and\ \citenamefont {Korobov}(2006)}]{tsogbayar2006relativistic}%
  \BibitemOpen
  \bibfield  {author} {\bibinfo {author} {\bibfnamefont {T.}~\bibnamefont {Tsogbayar}}\ and\ \bibinfo {author} {\bibfnamefont {V.}~\bibnamefont {Korobov}},\ }\bibfield  {title} {\bibinfo {title} {Relativistic correction to the 1s$\sigma$ and 2p$\sigma$ electronic states of the {H}$_2^+$ molecular ion and the moleculelike states of the antiprotonic helium {He}$^+$p{\={}}},\ }\href {https://doi.org/https://doi.org/10.1063/1.2209694} {\bibfield  {journal} {\bibinfo  {journal} {The Journal of Chemical Physics}\ }\textbf {\bibinfo {volume} {125}},\ \bibinfo {pages} {024308} (\bibinfo {year} {2006})}\BibitemShut {NoStop}%
\bibitem [{\citenamefont {Korobov}\ \emph {et~al.}(2020)\citenamefont {Korobov}, \citenamefont {Karr}, \citenamefont {Haidar},\ and\ \citenamefont {Zhong}}]{korobov2020hyperfine}%
  \BibitemOpen
  \bibfield  {author} {\bibinfo {author} {\bibfnamefont {V.~I.}\ \bibnamefont {Korobov}}, \bibinfo {author} {\bibfnamefont {J.-P.}\ \bibnamefont {Karr}}, \bibinfo {author} {\bibfnamefont {M.}~\bibnamefont {Haidar}},\ and\ \bibinfo {author} {\bibfnamefont {Z.-X.}\ \bibnamefont {Zhong}},\ }\bibfield  {title} {\bibinfo {title} {Hyperfine structure in the h 2+ and hd+ molecular ions at order m $\alpha$ 6},\ }\href {https://doi.org/10.1103/PhysRevA.102.022804} {\bibfield  {journal} {\bibinfo  {journal} {Physical Review A}\ }\textbf {\bibinfo {volume} {102}},\ \bibinfo {pages} {022804} (\bibinfo {year} {2020})}\BibitemShut {NoStop}%
\bibitem [{\citenamefont {McConnell}\ \emph {et~al.}(2012)\citenamefont {McConnell}, \citenamefont {Artemyev}, \citenamefont {Mai},\ and\ \citenamefont {Surzhykov}}]{PhysRevA.86.052705}%
  \BibitemOpen
  \bibfield  {author} {\bibinfo {author} {\bibfnamefont {S.~R.}\ \bibnamefont {McConnell}}, \bibinfo {author} {\bibfnamefont {A.~N.}\ \bibnamefont {Artemyev}}, \bibinfo {author} {\bibfnamefont {M.}~\bibnamefont {Mai}},\ and\ \bibinfo {author} {\bibfnamefont {A.}~\bibnamefont {Surzhykov}},\ }\bibfield  {title} {\bibinfo {title} {Solution of the two-center time-dependent dirac equation in spherical coordinates: Application of the multipole expansion of the electron-nuclei interaction},\ }\href {https://doi.org/10.1103/PhysRevA.86.052705} {\bibfield  {journal} {\bibinfo  {journal} {Phys. Rev. A}\ }\textbf {\bibinfo {volume} {86}},\ \bibinfo {pages} {052705} (\bibinfo {year} {2012})}\BibitemShut {NoStop}%
\bibitem [{\citenamefont {Tupitsyn}\ and\ \citenamefont {Mironova}(2014)}]{tupitsyn2014relativistic}%
  \BibitemOpen
  \bibfield  {author} {\bibinfo {author} {\bibfnamefont {I.}~\bibnamefont {Tupitsyn}}\ and\ \bibinfo {author} {\bibfnamefont {D.}~\bibnamefont {Mironova}},\ }\bibfield  {title} {\bibinfo {title} {Relativistic calculations of ground states of single-electron diatomic molecular ions},\ }\href {https://doi.org/https://doi.org/10.1134/S0030400X14090252} {\bibfield  {journal} {\bibinfo  {journal} {Optics and Spectroscopy}\ }\textbf {\bibinfo {volume} {117}},\ \bibinfo {pages} {351} (\bibinfo {year} {2014})}\BibitemShut {NoStop}%
\bibitem [{\citenamefont {Mironova}\ \emph {et~al.}(2015)\citenamefont {Mironova}, \citenamefont {Tupitsyn}, \citenamefont {Shabaev},\ and\ \citenamefont {Plunien}}]{mironova2015relativistic}%
  \BibitemOpen
  \bibfield  {author} {\bibinfo {author} {\bibfnamefont {D.}~\bibnamefont {Mironova}}, \bibinfo {author} {\bibfnamefont {I.}~\bibnamefont {Tupitsyn}}, \bibinfo {author} {\bibfnamefont {V.}~\bibnamefont {Shabaev}},\ and\ \bibinfo {author} {\bibfnamefont {G.}~\bibnamefont {Plunien}},\ }\bibfield  {title} {\bibinfo {title} {Relativistic calculations of the ground state energies and the critical distances for one-electron homonuclear quasi-molecules},\ }\href {https://doi.org/https://doi.org/10.1016/j.chemphys.2015.01.003} {\bibfield  {journal} {\bibinfo  {journal} {Chemical Physics}\ }\textbf {\bibinfo {volume} {449}},\ \bibinfo {pages} {10} (\bibinfo {year} {2015})}\BibitemShut {NoStop}%
\bibitem [{\citenamefont {Shabaev}\ \emph {et~al.}(2004)\citenamefont {Shabaev}, \citenamefont {Tupitsyn}, \citenamefont {Yerokhin}, \citenamefont {Plunien},\ and\ \citenamefont {Soff}}]{Shabaev_DKB}%
  \BibitemOpen
  \bibfield  {author} {\bibinfo {author} {\bibfnamefont {V.~M.}\ \bibnamefont {Shabaev}}, \bibinfo {author} {\bibfnamefont {I.~I.}\ \bibnamefont {Tupitsyn}}, \bibinfo {author} {\bibfnamefont {V.~A.}\ \bibnamefont {Yerokhin}}, \bibinfo {author} {\bibfnamefont {G.}~\bibnamefont {Plunien}},\ and\ \bibinfo {author} {\bibfnamefont {G.}~\bibnamefont {Soff}},\ }\bibfield  {title} {\bibinfo {title} {Dual kinetic balance approach to basis-set expansions for the dirac equation},\ }\href {https://doi.org/10.1103/PhysRevLett.93.130405} {\bibfield  {journal} {\bibinfo  {journal} {Phys. Rev. Lett.}\ }\textbf {\bibinfo {volume} {93}},\ \bibinfo {pages} {130405} (\bibinfo {year} {2004})}\BibitemShut {NoStop}%
\bibitem [{\citenamefont {Rozenbaum}\ \emph {et~al.}(2014)\citenamefont {Rozenbaum}, \citenamefont {Glazov}, \citenamefont {Shabaev}, \citenamefont {Sosnova},\ and\ \citenamefont {Telnov}}]{Rozenbaum_ADKB}%
  \BibitemOpen
  \bibfield  {author} {\bibinfo {author} {\bibfnamefont {E.~B.}\ \bibnamefont {Rozenbaum}}, \bibinfo {author} {\bibfnamefont {D.~A.}\ \bibnamefont {Glazov}}, \bibinfo {author} {\bibfnamefont {V.~M.}\ \bibnamefont {Shabaev}}, \bibinfo {author} {\bibfnamefont {K.~E.}\ \bibnamefont {Sosnova}},\ and\ \bibinfo {author} {\bibfnamefont {D.~A.}\ \bibnamefont {Telnov}},\ }\bibfield  {title} {\bibinfo {title} {Dual-kinetic-balance approach to the dirac equation for axially symmetric systems: Application to static and time-dependent fields},\ }\href {https://doi.org/10.1103/PhysRevA.89.012514} {\bibfield  {journal} {\bibinfo  {journal} {Phys. Rev. A}\ }\textbf {\bibinfo {volume} {89}},\ \bibinfo {pages} {012514} (\bibinfo {year} {2014})}\BibitemShut {NoStop}%
\bibitem [{\citenamefont {Anikin}\ \emph {et~al.}(2023)\citenamefont {Anikin}, \citenamefont {Danilov}, \citenamefont {Glazov}, \citenamefont {Kotov},\ and\ \citenamefont {Solovyev}}]{anikin2023light}%
  \BibitemOpen
  \bibfield  {author} {\bibinfo {author} {\bibfnamefont {A.}~\bibnamefont {Anikin}}, \bibinfo {author} {\bibfnamefont {A.}~\bibnamefont {Danilov}}, \bibinfo {author} {\bibfnamefont {D.}~\bibnamefont {Glazov}}, \bibinfo {author} {\bibfnamefont {A.}~\bibnamefont {Kotov}},\ and\ \bibinfo {author} {\bibfnamefont {D.}~\bibnamefont {Solovyev}},\ }\bibfield  {title} {\bibinfo {title} {Light antiproton one-electron quasi-molecular ions within the relativistic a-dkb method},\ }\bibfield  {journal} {\bibinfo  {journal} {The Journal of Chemical Physics}\ }\textbf {\bibinfo {volume} {159}},\ \href {https://doi.org/https://doi.org/10.1063/5.0181614} {https://doi.org/10.1063/5.0181614} (\bibinfo {year} {2023})\BibitemShut {NoStop}%
\bibitem [{\citenamefont {Danilov}\ \emph {et~al.}(2023)\citenamefont {Danilov}, \citenamefont {Anikin}, \citenamefont {Glazov}, \citenamefont {Korzinin}, \citenamefont {Kotov},\ and\ \citenamefont {Solovyev}}]{danilov2023}%
  \BibitemOpen
  \bibfield  {author} {\bibinfo {author} {\bibfnamefont {A.}~\bibnamefont {Danilov}}, \bibinfo {author} {\bibfnamefont {A.}~\bibnamefont {Anikin}}, \bibinfo {author} {\bibfnamefont {D.}~\bibnamefont {Glazov}}, \bibinfo {author} {\bibfnamefont {E.}~\bibnamefont {Korzinin}}, \bibinfo {author} {\bibfnamefont {A.}~\bibnamefont {Kotov}},\ and\ \bibinfo {author} {\bibfnamefont {D.}~\bibnamefont {Solovyev}},\ }\bibfield  {title} {\bibinfo {title} {Adiabaticheskiye potentsialy kvazimolekulyarnykh ionov $\mathrm{H}-p$, $\mathrm{He}^{+}-p$: relyativistskiy podkhod},\ }\href {https://journals.ioffe.ru/articles/viewPDF/57025} {\bibfield  {journal} {\bibinfo  {journal} {Optika i spektroskopiya}\ }\textbf {\bibinfo {volume} {131}} (\bibinfo {year} {2023})}\BibitemShut {NoStop}%
\bibitem [{\citenamefont {Solovyev}\ \emph {et~al.}(2024)\citenamefont {Solovyev}, \citenamefont {Anikin}, \citenamefont {Danilov}, \citenamefont {Glazov},\ and\ \citenamefont {Kotov}}]{Solovyev_2024}%
  \BibitemOpen
  \bibfield  {author} {\bibinfo {author} {\bibfnamefont {D.}~\bibnamefont {Solovyev}}, \bibinfo {author} {\bibfnamefont {A.}~\bibnamefont {Anikin}}, \bibinfo {author} {\bibfnamefont {A.}~\bibnamefont {Danilov}}, \bibinfo {author} {\bibfnamefont {D.}~\bibnamefont {Glazov}},\ and\ \bibinfo {author} {\bibfnamefont {A.}~\bibnamefont {Kotov}},\ }\bibfield  {title} {\bibinfo {title} {Light one-electron molecular ions within the finite-basis-set method for the two-center dirac equation},\ }\href {https://doi.org/10.1088/1402-4896/ad2e66} {\bibfield  {journal} {\bibinfo  {journal} {Physica Scripta}\ }\textbf {\bibinfo {volume} {99}},\ \bibinfo {pages} {045401} (\bibinfo {year} {2024})}\BibitemShut {NoStop}%
\bibitem [{\citenamefont {Beiersdorfer}\ \emph {et~al.}(1999)\citenamefont {Beiersdorfer}, \citenamefont {Schweikhard}, \citenamefont {Olson}, \citenamefont {Brown}, \citenamefont {Utter}, \citenamefont {L{\'o}pez-Urrutia},\ and\ \citenamefont {Widmann}}]{beiersdorfer1999x}%
  \BibitemOpen
  \bibfield  {author} {\bibinfo {author} {\bibfnamefont {P.}~\bibnamefont {Beiersdorfer}}, \bibinfo {author} {\bibfnamefont {L.}~\bibnamefont {Schweikhard}}, \bibinfo {author} {\bibfnamefont {R.}~\bibnamefont {Olson}}, \bibinfo {author} {\bibfnamefont {G.}~\bibnamefont {Brown}}, \bibinfo {author} {\bibfnamefont {S.}~\bibnamefont {Utter}}, \bibinfo {author} {\bibfnamefont {J.~C.}\ \bibnamefont {L{\'o}pez-Urrutia}},\ and\ \bibinfo {author} {\bibfnamefont {K.}~\bibnamefont {Widmann}},\ }\bibfield  {title} {\bibinfo {title} {X-ray measurements of charge transfer reactions involving cold, very highly charged ions},\ }\href {https://dx.doi.org/10.1238/Physica.Topical.080a00121} {\bibfield  {journal} {\bibinfo  {journal} {Physica Scripta}\ }\textbf {\bibinfo {volume} {1999}},\ \bibinfo {pages} {121} (\bibinfo {year} {1999})}\BibitemShut {NoStop}%
\bibitem [{\citenamefont {Kasahara}\ \emph {et~al.}(2017)\citenamefont {Kasahara}, \citenamefont {Takeuchi}, \citenamefont {Zadik}, \citenamefont {Takabayashi}, \citenamefont {Colman}, \citenamefont {McDonald}, \citenamefont {Rosseinsky}, \citenamefont {Prassides},\ and\ \citenamefont {Iwasa}}]{kasahara2017upper}%
  \BibitemOpen
  \bibfield  {author} {\bibinfo {author} {\bibfnamefont {Y.}~\bibnamefont {Kasahara}}, \bibinfo {author} {\bibfnamefont {Y.}~\bibnamefont {Takeuchi}}, \bibinfo {author} {\bibfnamefont {R.}~\bibnamefont {Zadik}}, \bibinfo {author} {\bibfnamefont {Y.}~\bibnamefont {Takabayashi}}, \bibinfo {author} {\bibfnamefont {R.}~\bibnamefont {Colman}}, \bibinfo {author} {\bibfnamefont {R.~D.}\ \bibnamefont {McDonald}}, \bibinfo {author} {\bibfnamefont {M.}~\bibnamefont {Rosseinsky}}, \bibinfo {author} {\bibfnamefont {K.}~\bibnamefont {Prassides}},\ and\ \bibinfo {author} {\bibfnamefont {Y.}~\bibnamefont {Iwasa}},\ }\bibfield  {title} {\bibinfo {title} {Upper critical field reaches 90 tesla near the mott transition in fulleride superconductors},\ }\href {https://doi.org/https://doi.org/10.1038/ncomms14467} {\bibfield  {journal} {\bibinfo  {journal} {Nature communications}\ }\textbf {\bibinfo {volume} {8}},\ \bibinfo {pages} {14467} (\bibinfo {year} {2017})}\BibitemShut {NoStop}%
\bibitem [{\citenamefont {Modic}\ \emph {et~al.}(2017)\citenamefont {Modic}, \citenamefont {Ramshaw}, \citenamefont {Betts}, \citenamefont {Breznay}, \citenamefont {Analytis}, \citenamefont {McDonald},\ and\ \citenamefont {Shekhter}}]{modic2017robust}%
  \BibitemOpen
  \bibfield  {author} {\bibinfo {author} {\bibfnamefont {K.~A.}\ \bibnamefont {Modic}}, \bibinfo {author} {\bibfnamefont {B.~J.}\ \bibnamefont {Ramshaw}}, \bibinfo {author} {\bibfnamefont {J.}~\bibnamefont {Betts}}, \bibinfo {author} {\bibfnamefont {N.~P.}\ \bibnamefont {Breznay}}, \bibinfo {author} {\bibfnamefont {J.~G.}\ \bibnamefont {Analytis}}, \bibinfo {author} {\bibfnamefont {R.~D.}\ \bibnamefont {McDonald}},\ and\ \bibinfo {author} {\bibfnamefont {A.}~\bibnamefont {Shekhter}},\ }\bibfield  {title} {\bibinfo {title} {Robust spin correlations at high magnetic fields in the harmonic honeycomb iridates},\ }\href {https://doi.org/https://doi.org/10.1038/s41467-017-00264-6} {\bibfield  {journal} {\bibinfo  {journal} {Nature communications}\ }\textbf {\bibinfo {volume} {8}},\ \bibinfo {pages} {180} (\bibinfo {year} {2017})}\BibitemShut {NoStop}%
\bibitem [{\citenamefont {Wille}(1988)}]{wille1988magnetically}%
  \BibitemOpen
  \bibfield  {author} {\bibinfo {author} {\bibfnamefont {U.}~\bibnamefont {Wille}},\ }\bibfield  {title} {\bibinfo {title} {Magnetically dressed one-electron molecular orbitals},\ }\href {https://doi.org/https://doi.org/10.1103/PhysRevA.38.3210} {\bibfield  {journal} {\bibinfo  {journal} {Physical Review A}\ }\textbf {\bibinfo {volume} {38}},\ \bibinfo {pages} {3210} (\bibinfo {year} {1988})}\BibitemShut {NoStop}%
\bibitem [{\citenamefont {Turbiner}\ and\ \citenamefont {Vieyra}(2004)}]{turbiner2004h}%
  \BibitemOpen
  \bibfield  {author} {\bibinfo {author} {\bibfnamefont {A.~V.}\ \bibnamefont {Turbiner}}\ and\ \bibinfo {author} {\bibfnamefont {J.~L.}\ \bibnamefont {Vieyra}},\ }\bibfield  {title} {\bibinfo {title} {H 2+ ion in a strong magnetic field: Lowest excited states},\ }\href {https://doi.org/https://doi.org/10.1103/PhysRevA.69.053413} {\bibfield  {journal} {\bibinfo  {journal} {Physical Review A}\ }\textbf {\bibinfo {volume} {69}},\ \bibinfo {pages} {053413} (\bibinfo {year} {2004})}\BibitemShut {NoStop}%
\bibitem [{\citenamefont {Olivares-Pil{\'o}n}\ \emph {et~al.}(2010)\citenamefont {Olivares-Pil{\'o}n}, \citenamefont {Baye}, \citenamefont {Turbiner},\ and\ \citenamefont {Vieyra}}]{olivares2010one}%
  \BibitemOpen
  \bibfield  {author} {\bibinfo {author} {\bibfnamefont {H.}~\bibnamefont {Olivares-Pil{\'o}n}}, \bibinfo {author} {\bibfnamefont {D.}~\bibnamefont {Baye}}, \bibinfo {author} {\bibfnamefont {A.}~\bibnamefont {Turbiner}},\ and\ \bibinfo {author} {\bibfnamefont {J.~L.}\ \bibnamefont {Vieyra}},\ }\bibfield  {title} {\bibinfo {title} {One-electron atomic--molecular ions containing lithium in a strong magnetic field},\ }\href {https://dx.doi.org/10.1088/0953-4075/43/6/065702} {\bibfield  {journal} {\bibinfo  {journal} {Journal of Physics B: Atomic, Molecular and Optical Physics}\ }\textbf {\bibinfo {volume} {43}},\ \bibinfo {pages} {065702} (\bibinfo {year} {2010})}\BibitemShut {NoStop}%
\bibitem [{\citenamefont {Schmelcher}\ and\ \citenamefont {Cederbaum}(1997)}]{schmelcher1997molecules}%
  \BibitemOpen
  \bibfield  {author} {\bibinfo {author} {\bibfnamefont {P.}~\bibnamefont {Schmelcher}}\ and\ \bibinfo {author} {\bibfnamefont {L.}~\bibnamefont {Cederbaum}},\ }\bibfield  {title} {\bibinfo {title} {Molecules in strong magnetic fields: Some perspectives and general aspects},\ }\href {https://doi.org/https://doi.org/10.1002/(SICI)1097-461X(1997)64:5<501::AID-QUA3>3.0.CO;2-\#} {\bibfield  {journal} {\bibinfo  {journal} {International journal of quantum chemistry}\ }\textbf {\bibinfo {volume} {64}},\ \bibinfo {pages} {501} (\bibinfo {year} {1997})}\BibitemShut {NoStop}%
\bibitem [{\citenamefont {Lehtola}\ \emph {et~al.}(2020)\citenamefont {Lehtola}, \citenamefont {Dimitrova},\ and\ \citenamefont {Sundholm}}]{lehtola2020fully}%
  \BibitemOpen
  \bibfield  {author} {\bibinfo {author} {\bibfnamefont {S.}~\bibnamefont {Lehtola}}, \bibinfo {author} {\bibfnamefont {M.}~\bibnamefont {Dimitrova}},\ and\ \bibinfo {author} {\bibfnamefont {D.}~\bibnamefont {Sundholm}},\ }\bibfield  {title} {\bibinfo {title} {Fully numerical electronic structure calculations on diatomic molecules in weak to strong magnetic fields},\ }\href {https://doi.org/https://doi.org/10.1080/00268976.2019.1597989} {\bibfield  {journal} {\bibinfo  {journal} {Molecular Physics}\ }\textbf {\bibinfo {volume} {118}},\ \bibinfo {pages} {e1597989} (\bibinfo {year} {2020})}\BibitemShut {NoStop}%
\bibitem [{\citenamefont {Schmelcher}\ \emph {et~al.}(1988)\citenamefont {Schmelcher}, \citenamefont {Cederbaum},\ and\ \citenamefont {Meyer}}]{schmelcher1988electronic}%
  \BibitemOpen
  \bibfield  {author} {\bibinfo {author} {\bibfnamefont {P.}~\bibnamefont {Schmelcher}}, \bibinfo {author} {\bibfnamefont {L.}~\bibnamefont {Cederbaum}},\ and\ \bibinfo {author} {\bibfnamefont {H.-D.}\ \bibnamefont {Meyer}},\ }\bibfield  {title} {\bibinfo {title} {Electronic and nuclear motion and their couplings in the presence of a magnetic field},\ }\href {https://doi.org/https://doi.org/10.1103/PhysRevA.38.6066} {\bibfield  {journal} {\bibinfo  {journal} {Physical Review A}\ }\textbf {\bibinfo {volume} {38}},\ \bibinfo {pages} {6066} (\bibinfo {year} {1988})}\BibitemShut {NoStop}%
\bibitem [{\citenamefont {Johnson}\ \emph {et~al.}(1988)\citenamefont {Johnson}, \citenamefont {Blundell},\ and\ \citenamefont {Sapirstein}}]{Johnson_Bspline}%
  \BibitemOpen
  \bibfield  {author} {\bibinfo {author} {\bibfnamefont {W.~R.}\ \bibnamefont {Johnson}}, \bibinfo {author} {\bibfnamefont {S.~A.}\ \bibnamefont {Blundell}},\ and\ \bibinfo {author} {\bibfnamefont {J.}~\bibnamefont {Sapirstein}},\ }\bibfield  {title} {\bibinfo {title} {Finite basis sets for the {D}irac equation constructed from {B} splines},\ }\href {https://doi.org/10.1103/PhysRevA.37.307} {\bibfield  {journal} {\bibinfo  {journal} {Phys. Rev. A}\ }\textbf {\bibinfo {volume} {37}},\ \bibinfo {pages} {307} (\bibinfo {year} {1988})}\BibitemShut {NoStop}%
\bibitem [{\citenamefont {Sapirstein}\ and\ \citenamefont {Johnson}(1996)}]{Sapirstein_1996}%
  \BibitemOpen
  \bibfield  {author} {\bibinfo {author} {\bibfnamefont {J.}~\bibnamefont {Sapirstein}}\ and\ \bibinfo {author} {\bibfnamefont {W.~R.}\ \bibnamefont {Johnson}},\ }\bibfield  {title} {\bibinfo {title} {The use of basis splines in theoretical atomic physics},\ }\href {https://doi.org/10.1088/0953-4075/29/22/005} {\bibfield  {journal} {\bibinfo  {journal} {Journal of Physics B: Atomic, Molecular and Optical Physics}\ }\textbf {\bibinfo {volume} {29}},\ \bibinfo {pages} {5213} (\bibinfo {year} {1996})}\BibitemShut {NoStop}%
\bibitem [{\citenamefont {L{\'o}pez-Urrutia}\ \emph {et~al.}(1996)\citenamefont {L{\'o}pez-Urrutia}, \citenamefont {Beiersdorfer}, \citenamefont {Savin},\ and\ \citenamefont {Widmann}}]{lopez1996direct}%
  \BibitemOpen
  \bibfield  {author} {\bibinfo {author} {\bibfnamefont {J.~R.~C.}\ \bibnamefont {L{\'o}pez-Urrutia}}, \bibinfo {author} {\bibfnamefont {P.}~\bibnamefont {Beiersdorfer}}, \bibinfo {author} {\bibfnamefont {D.~W.}\ \bibnamefont {Savin}},\ and\ \bibinfo {author} {\bibfnamefont {K.}~\bibnamefont {Widmann}},\ }\bibfield  {title} {\bibinfo {title} {Direct observation of the spontaneous emission of the hyperfine transition f= 4 to f= 3 in ground state hydrogenlike h 165 o 6 6+ in an electron beam ion trap},\ }\href {https://doi.org/https://doi.org/10.1103/PhysRevLett.77.826} {\bibfield  {journal} {\bibinfo  {journal} {Physical review letters}\ }\textbf {\bibinfo {volume} {77}},\ \bibinfo {pages} {826} (\bibinfo {year} {1996})}\BibitemShut {NoStop}%
\bibitem [{\citenamefont {L{\'o}pez-Urrutia}\ \emph {et~al.}(1998)\citenamefont {L{\'o}pez-Urrutia}, \citenamefont {Beiersdorfer}, \citenamefont {Widmann}, \citenamefont {Birkett}, \citenamefont {M{\aa}rtensson-Pendrill},\ and\ \citenamefont {Gustavsson}}]{lopez1998nuclear}%
  \BibitemOpen
  \bibfield  {author} {\bibinfo {author} {\bibfnamefont {J.~C.}\ \bibnamefont {L{\'o}pez-Urrutia}}, \bibinfo {author} {\bibfnamefont {P.}~\bibnamefont {Beiersdorfer}}, \bibinfo {author} {\bibfnamefont {K.}~\bibnamefont {Widmann}}, \bibinfo {author} {\bibfnamefont {B.}~\bibnamefont {Birkett}}, \bibinfo {author} {\bibfnamefont {A.-M.}\ \bibnamefont {M{\aa}rtensson-Pendrill}},\ and\ \bibinfo {author} {\bibfnamefont {M.}~\bibnamefont {Gustavsson}},\ }\bibfield  {title} {\bibinfo {title} {Nuclear magnetization distribution radii determined by hyperfine transitions in the 1 s level of h-like ions 185 re 7 4+ and 187 re 7 4+},\ }\href {https://doi.org/https://doi.org/10.1103/PhysRevA.57.879} {\bibfield  {journal} {\bibinfo  {journal} {Physical Review A}\ }\textbf {\bibinfo {volume} {57}},\ \bibinfo {pages} {879} (\bibinfo {year} {1998})}\BibitemShut {NoStop}%
\bibitem [{\citenamefont {Seelig}\ \emph {et~al.}(1998)\citenamefont {Seelig}, \citenamefont {Borneis}, \citenamefont {Dax}, \citenamefont {Engel}, \citenamefont {Faber}, \citenamefont {Gerlach}, \citenamefont {Holbrow}, \citenamefont {Huber}, \citenamefont {K{\"u}hl}, \citenamefont {Marx} \emph {et~al.}}]{seelig1998ground}%
  \BibitemOpen
  \bibfield  {author} {\bibinfo {author} {\bibfnamefont {P.}~\bibnamefont {Seelig}}, \bibinfo {author} {\bibfnamefont {S.}~\bibnamefont {Borneis}}, \bibinfo {author} {\bibfnamefont {A.}~\bibnamefont {Dax}}, \bibinfo {author} {\bibfnamefont {T.}~\bibnamefont {Engel}}, \bibinfo {author} {\bibfnamefont {S.}~\bibnamefont {Faber}}, \bibinfo {author} {\bibfnamefont {M.}~\bibnamefont {Gerlach}}, \bibinfo {author} {\bibfnamefont {C.}~\bibnamefont {Holbrow}}, \bibinfo {author} {\bibfnamefont {G.}~\bibnamefont {Huber}}, \bibinfo {author} {\bibfnamefont {T.}~\bibnamefont {K{\"u}hl}}, \bibinfo {author} {\bibfnamefont {D.}~\bibnamefont {Marx}}, \emph {et~al.},\ }\bibfield  {title} {\bibinfo {title} {Ground state hyperfine splitting of hydrogenlike 207 pb 8 1+ by laser excitation of a bunched ion beam in the gsi experimental storage ring},\ }\href {https://doi.org/https://doi.org/10.1103/PhysRevLett.81.4824} {\bibfield  {journal} {\bibinfo  {journal} {Physical review letters}\ }\textbf {\bibinfo {volume} {81}},\ \bibinfo
  {pages} {4824} (\bibinfo {year} {1998})}\BibitemShut {NoStop}%
\bibitem [{\citenamefont {Georgiadis}\ \emph {et~al.}(1986)\citenamefont {Georgiadis}, \citenamefont {M{\"u}ller}, \citenamefont {Str{\"a}ter}, \citenamefont {Gassen}, \citenamefont {Von~Brentano}, \citenamefont {Sens},\ and\ \citenamefont {Pape}}]{georgiadis1986measurement}%
  \BibitemOpen
  \bibfield  {author} {\bibinfo {author} {\bibfnamefont {A.}~\bibnamefont {Georgiadis}}, \bibinfo {author} {\bibfnamefont {D.}~\bibnamefont {M{\"u}ller}}, \bibinfo {author} {\bibfnamefont {H.-D.}\ \bibnamefont {Str{\"a}ter}}, \bibinfo {author} {\bibfnamefont {J.}~\bibnamefont {Gassen}}, \bibinfo {author} {\bibfnamefont {P.}~\bibnamefont {Von~Brentano}}, \bibinfo {author} {\bibfnamefont {J.}~\bibnamefont {Sens}},\ and\ \bibinfo {author} {\bibfnamefont {A.}~\bibnamefont {Pape}},\ }\bibfield  {title} {\bibinfo {title} {Measurement of the lamb shift in hydrogenic sulfur by laser spectroscopy},\ }\href {https://doi.org/https://doi.org/10.1016/0375-9601(86)90033-2} {\bibfield  {journal} {\bibinfo  {journal} {Physics Letters A}\ }\textbf {\bibinfo {volume} {115}},\ \bibinfo {pages} {108} (\bibinfo {year} {1986})}\BibitemShut {NoStop}%
\bibitem [{\citenamefont {Angeli}\ and\ \citenamefont {Marinova}(2013)}]{angeli2013table}%
  \BibitemOpen
  \bibfield  {author} {\bibinfo {author} {\bibfnamefont {I.}~\bibnamefont {Angeli}}\ and\ \bibinfo {author} {\bibfnamefont {K.}~\bibnamefont {Marinova}},\ }\bibfield  {title} {\bibinfo {title} {Table of experimental nuclear ground state charge radii: An update},\ }\href {https://doi.org/https://doi.org/10.1016/j.adt.2011.12.006} {\bibfield  {journal} {\bibinfo  {journal} {Atomic Data and Nuclear Data Tables}\ }\textbf {\bibinfo {volume} {99}},\ \bibinfo {pages} {69} (\bibinfo {year} {2013})}\BibitemShut {NoStop}%
\bibitem [{\citenamefont {Tiesinga}\ \emph {et~al.}(2021)\citenamefont {Tiesinga}, \citenamefont {Mohr}, \citenamefont {Newell},\ and\ \citenamefont {Taylor}}]{RevModPhys.93.025010}%
  \BibitemOpen
  \bibfield  {author} {\bibinfo {author} {\bibfnamefont {E.}~\bibnamefont {Tiesinga}}, \bibinfo {author} {\bibfnamefont {P.~J.}\ \bibnamefont {Mohr}}, \bibinfo {author} {\bibfnamefont {D.~B.}\ \bibnamefont {Newell}},\ and\ \bibinfo {author} {\bibfnamefont {B.~N.}\ \bibnamefont {Taylor}},\ }\bibfield  {title} {\bibinfo {title} {Codata recommended values of the fundamental physical constants: 2018},\ }\href {https://doi.org/10.1103/RevModPhys.93.025010} {\bibfield  {journal} {\bibinfo  {journal} {Rev. Mod. Phys.}\ }\textbf {\bibinfo {volume} {93}},\ \bibinfo {pages} {025010} (\bibinfo {year} {2021})}\BibitemShut {NoStop}%
\bibitem [{\citenamefont {Sommerfeldt}\ \emph {et~al.}(2020)\citenamefont {Sommerfeldt}, \citenamefont {M\"uller}, \citenamefont {Volotka}, \citenamefont {Fritzsche},\ and\ \citenamefont {Surzhykov}}]{PhysRevA.102.042811}%
  \BibitemOpen
  \bibfield  {author} {\bibinfo {author} {\bibfnamefont {J.}~\bibnamefont {Sommerfeldt}}, \bibinfo {author} {\bibfnamefont {R.~A.}\ \bibnamefont {M\"uller}}, \bibinfo {author} {\bibfnamefont {A.~V.}\ \bibnamefont {Volotka}}, \bibinfo {author} {\bibfnamefont {S.}~\bibnamefont {Fritzsche}},\ and\ \bibinfo {author} {\bibfnamefont {A.}~\bibnamefont {Surzhykov}},\ }\bibfield  {title} {\bibinfo {title} {Vacuum polarization and finite-nuclear-size effects in the two-photon decay of hydrogenlike ions},\ }\href {https://doi.org/10.1103/PhysRevA.102.042811} {\bibfield  {journal} {\bibinfo  {journal} {Phys. Rev. A}\ }\textbf {\bibinfo {volume} {102}},\ \bibinfo {pages} {042811} (\bibinfo {year} {2020})}\BibitemShut {NoStop}%
\bibitem [{\citenamefont {{Genkin, M.}}\ and\ \citenamefont {{Lindroth, E.}}(2009)}]{Genkin_2009}%
  \BibitemOpen
  \bibfield  {author} {\bibinfo {author} {\bibnamefont {{Genkin, M.}}}\ and\ \bibinfo {author} {\bibnamefont {{Lindroth, E.}}},\ }\bibfield  {title} {\bibinfo {title} {Possibility of resonant capture of antiprotons by highly charged hydrogenlike ions},\ }\href {https://doi.org/10.1140/epjd/e2008-00273-1} {\bibfield  {journal} {\bibinfo  {journal} {Eur. Phys. J. D}\ }\textbf {\bibinfo {volume} {51}},\ \bibinfo {pages} {205} (\bibinfo {year} {2009})}\BibitemShut {NoStop}%
\bibitem [{\citenamefont {Doser}(2022)}]{Doser_2022}%
  \BibitemOpen
  \bibfield  {author} {\bibinfo {author} {\bibfnamefont {M.}~\bibnamefont {Doser}},\ }\bibfield  {title} {\bibinfo {title} {Antiprotonic bound systems},\ }\href {https://doi.org/https://doi.org/10.1016/j.ppnp.2022.103964} {\bibfield  {journal} {\bibinfo  {journal} {Progress in Particle and Nuclear Physics}\ }\textbf {\bibinfo {volume} {125}},\ \bibinfo {pages} {103964} (\bibinfo {year} {2022})}\BibitemShut {NoStop}%
\end{thebibliography}%

\begin{table}[ht]
\centering
\caption{The values of the Zeeman shift $\Delta E_{\mathrm{Z}}$ (in a.u.) for the ground state of the compound $\mathrm{He}^{+} -\bar{p}$ at a magnetic field strength of $100$ T are indicated in the third column. The first column shows the values of the inter-nuclear distances (in fermi, fm). The second column shows the energies of electrons in the absence of a magnetic field in atomic units. The values of the Zeeman shift at a different magnetic field strength are not given due to their independence from the inter-nuclear distance.}
\label{tab:2}
\begin{tabular}{ c c c }
    \hline
    \hline
        $R$, fm & $E_{\mathrm{A-DKB}}$, a.u. & $ | \Delta E_{\mathrm{Z}} | \times 10^6 $, au \\ \noalign{\smallskip}
        \hline
        $0.0$ & $-0.50000665$ & $212.76$ \\ 
        $5291.77249$ & $-0.51232902$ & $212.76$ \\ 
        $10583.54498$ & $-0.54658328$ & $212.75$ \\ 
        $15875.31747$ & $-0.59948446$ & $212.75$ \\ 
        $21167.08996$ & $-0.66690910$ & $212.74$ \\ 
        $26458.86245$ & $-0.74359023$ & $212.73$ \\ 
        $31750.63494$ & $-0.82415224$ & $212.73$ \\ 
        $37042.40743$ & $-0.90419211$ & $212.73$ \\ 
        $42334.17992$ & $-0.98072434$ & $212.72$ \\ 
        $47625.95241$ & $-1.05206955$ & $212.72$ \\ 
        $52917.72490$ & $-1.11751295$ & $212.72$ \\ 
        $58209.49739$ & $-1.17695850$ & $212.72$ \\ 
        $63501.26988$ & $-1.23066400$ & $212.72$ \\ 
        $68793.04237$ & $-1.27906400$ & $212.72$ \\ 
        $74084.81486$ & $-1.32266143$ & $212.72$ \\ 
        $79376.58735$ & $-1.36196558$ & $212.72$ \\ 
        $84668.35984$ & $-1.39745954$ & $212.72$ \\ 
        $89960.13233$ & $-1.42958499$ & $212.72$ \\ 
        $95251.90482$ & $-1.45873706$ & $212.72$ \\ 
        $100543.67731$ & $-1.48526460$ & $212.72$ \\ 
        $105835.44980$ & $-1.50947315$ & $212.72$ \\ 
    \hline
    \hline
    \end{tabular}
\end{table}
\begin{table}[ht]
\centering
\caption{The same as in Table~\ref{tab:2} for the $\mathrm{Li}^{2+} - \bar{p}$ compound.}
\label{tab:3}
\begin{tabular}{ c c c }
    \hline
    \hline
        $R$, fm & $E_{\mathrm{A-DKB}}$, a.u. & $ | \Delta E_{\mathrm{Z}} | \times 10^6 $, au \\ \noalign{\smallskip}
        \hline
        $0.0$ & $-2.00010648$ & $212.71$ \\ 
        $3527.84833$ & $-2.03167805$ & $212.71$ \\ 
        $7055.69665$ & $-2.11360541$ & $212.71$ \\ 
        $10583.54498$ & $-2.22996804$ & $212.71$ \\ 
        $14111.39331$ & $-2.36677964$ & $212.71$ \\ 
        $17639.24163$ & $-2.51210687$ & $212.70$ \\ 
        $21167.08996$ & $-2.65691048$ & $212.70$ \\ 
        $24694.93829$ & $-2.79524842$ & $212.70$ \\ 
        $28222.78661$ & $-2.92381308$ & $212.69$ \\ 
        $31750.63494$ & $-3.04120258$ & $212.69$ \\ 
        $35278.48327$ & $-3.14723436$ & $212.69$ \\ 
        $38806.33159$ & $-3.24242367$ & $212.69$ \\ 
        $42334.17992$ & $-3.32763343$ & $212.69$ \\ 
        $45862.02825$ & $-3.40385842$ & $212.69$ \\ 
        $49389.87657$ & $-3.47210176$ & $212.69$ \\ 
        $52917.72490$ & $-3.53330998$ & $212.69$ \\ 
        $56445.57323$ & $-3.58834337$ & $212.69$ \\ 
        $59973.42155$ & $-3.63796645$ & $212.69$ \\ 
        $63501.26988$ & $-3.68284942$ & $212.69$ \\ 
        $67029.11821$ & $-3.72357495$ & $212.69$ \\ 
        $70556.96653$ & $-3.76064721$ & $212.69$ \\ 
    \hline
    \hline
    \end{tabular}
\end{table}
\begin{table}[ht]
\centering
\caption{The same as in Table~\ref{tab:2} for the $\mathrm{C}^{5+} - \bar{p}$ compound.}
\label{tab:4}
\begin{tabular}{ c c c }
    \hline
    \hline
        $R$, fm & $E_{\mathrm{A-DKB}}$, a.u. & $ | \Delta E_{\mathrm{Z}} | \times 10^6 $, au \\ \noalign{\smallskip}
        \hline
        $0.0$ & $-12.50416195$ & $212.63$ \\ 
        $1763.92416$ & $-12.59936726$ & $212.62$ \\ 
        $3527.84833$ & $-12.83172810$ & $212.62$ \\ 
        $5291.77249$ & $-13.14049340$ & $212.61$ \\ 
        $7055.69665$ & $-13.48287107$ & $212.61$ \\ 
        $8819.62082$ & $-13.82992424$ & $212.60$ \\ 
        $10583.54498$ & $-14.16368683$ & $212.60$ \\ 
        $12347.46914$ & $-14.47431391$ & $212.59$ \\ 
        $14111.39331$ & $-14.75748499$ & $212.59$ \\ 
        $15875.31747$ & $-15.01233849$ & $212.59$ \\ 
        $17639.24163$ & $-15.23999327$ & $212.59$ \\ 
        $19403.16580$ & $-15.44257313$ & $212.59$ \\ 
        $21167.08996$ & $-15.62260484$ & $212.59$ \\ 
        $22931.01412$ & $-15.78266972$ & $212.59$ \\ 
        $24694.93829$ & $-15.92521715$ & $212.59$ \\ 
        $26458.86245$ & $-16.05247671$ & $212.58$ \\ 
        $28222.78661$ & $-16.16642752$ & $212.58$ \\ 
        $29986.71078$ & $-16.26879914$ & $212.58$ \\ 
        $31750.63494$ & $-16.36108853$ & $212.58$ \\ 
        $33514.55910$ & $-16.44458386$ & $212.58$ \\ 
        $35278.48327$ & $-16.52039028$ & $212.58$ \\
    \hline
    \hline
    \end{tabular}
\end{table}
\begin{table}[ht]
\centering
\caption{The same as in Table~\ref{tab:2} for the $\mathrm{S}^{15+} - \bar{p}$ compound. For this quasi-molecule, the values of the Zeeman shift in the $10$ T and $100$ T fields are indicated in the third and fourth columns, respectively.}
\label{tab:5}
\begin{tabular}{ c c c c }
    \hline
    \hline
        \multirow{2}{*}{$R$, fm} & \multicolumn{1}{c|}{\multirow{2}{*}{$E_{\mathrm{A-DKB}}$, a.u.}} & \multicolumn{2}{c}{$ | \Delta E_{\mathrm{Z}} | \times 10^6 $, a.u.} \\ \cline{3-4} 
                       & \multicolumn{1}{c|}{}                                  & $B = 10$ T                    & $B = 100$ T                   \\ \hline
        $0.0$ & $-112.83886661$ & $21.19$ & $211.87$ \\ 
        $661.47156$ & $-113.16215620$ & $21.19$ & $211.86$ \\ 
        $1322.94312$ & $-113.91709032$ & $21.18$ & $211.84$ \\ 
        $1984.41468$ & $-114.88068920$ & $21.18$ & $211.83$ \\ 
        $2645.88624$ & $-115.91503204$ & $21.18$ & $211.81$ \\ 
        $3307.3578$ & $-116.93787284$ & $21.18$ & $211.80$ \\ 
        $3968.82936$ & $-117.90372807$ & $21.18$ & $211.78$ \\ 
        $4630.30092$ & $-118.79068655$ & $21.18$ & $211.77$ \\ 
        $5291.77249$ & $-119.59129326$ & $21.18$ & $211.77$ \\ 
        $5953.24405$ & $-120.30645964$ & $21.18$ & $211.76$ \\ 
        $6614.71561$ & $-120.94155048$ & $21.18$ & $211.76$ \\ 
        $7276.18717$ & $-121.50397157$ & $21.17$ & $211.75$ \\ 
        $7937.65873$ & $-122.00174664$ & $21.17$ & $211.75$ \\ 
        $8599.13029$ & $-122.44272501$ & $21.17$ & $211.75$ \\ 
        $9260.60185$ & $-122.83417789$ & $21.17$ & $211.75$ \\ 
        $9922.07341$ & $-123.18262617$ & $21.17$ & $211.75$ \\ 
        $10583.54498$ & $-123.49380039$ & $21.17$ & $211.75$ \\ 
        $11245.01654$ & $-123.77267191$ & $21.17$ & $211.75$ \\ 
        $11906.48810$ & $-124.02351854$ & $21.17$ & $211.75$ \\ 
        $12567.95966$ & $-124.25000340$ & $21.17$ & $211.75$ \\ 
        $13229.43122$ & $-124.45525546$ & $21.17$ & $211.75$ \\
    \hline
    \hline
    \end{tabular}
\end{table}
\begin{table}[ht]
\centering
\caption{The same as in Table~\ref{tab:5} for the $\mathrm{Kr}^{35+} - \bar{p}$ compound.}
\label{tab:6}
\begin{tabular}{ c c c c }
    \hline
    \hline
        \multirow{2}{*}{$R$, fm} & \multicolumn{1}{c|}{\multirow{2}{*}{$E_{\mathrm{ADKB}}$, a.u.}} & \multicolumn{2}{c}{$ | \Delta E_{\mathrm{Z}} | \times 10^6 $} \\ \cline{3-4} 
                       & \multicolumn{1}{c|}{}                                  & $B = 10$ T                    & $B = 100$ T                   \\ \hline
        $0.0$ & $-622.81876775$ & $20.80$ & $208.02$ \\ 
        $293.98736$ & $-623.71972536$ & $20.80$ & $208.00$ \\ 
        $587.97472$ & $-625.69949444$ & $20.80$ & $207.96$ \\ 
        $881.96208$ & $-628.13546078$ & $20.79$ & $207.91$ \\ 
        $1175.94944$ & $-630.68467991$ & $20.79$ & $207.87$ \\ 
        $1469.93680$ & $-633.15910731$ & $20.78$ & $207.84$ \\ 
        $1763.92416$ & $-635.46389756$ & $20.78$ & $207.81$ \\ 
        $2057.91152$ & $-637.55844470$ & $20.78$ & $207.79$ \\ 
        $2351.89888$ & $-639.43357940$ & $20.77$ & $207.78$ \\ 
        $2645.88624$ & $-641.09748949$ & $20.77$ & $207.76$ \\ 
        $2939.87361$ & $-642.56707902$ & $20.77$ & $207.76$ \\ 
        $3233.86097$ & $-643.86273369$ & $20.75$ & $207.75$ \\ 
        $3527.84833$ & $-645.00528107$ & $20.74$ & $207.74$ \\ 
        $3821.83569$ & $-646.01440491$ & $20.74$ & $207.74$ \\ 
        $4115.82308$ & $-646.90793501$ & $20.74$ & $207.74$ \\ 
        $4409.81041$ & $-647.70161658$ & $20.74$ & $207.74$ \\ 
        $4703.79777$ & $-648.40913245$ & $20.74$ & $207.74$ \\ 
        $4997.78513$ & $-649.04225174$ & $20.73$ & $207.74$ \\ 
        $5291.77249$ & $-649.61103315$ & $20.74$ & $207.74$ \\ 
        $5585.75985$ & $-650.12404223$ & $20.74$ & $207.74$ \\ 
        $5879.74721$ & $-650.58856102$ & $20.74$ & $207.74$ \\ 
    \hline
    \hline
    \end{tabular}
\end{table}
%\resizebox{\linewidth}{!}{
\begin{table}[ht]
\centering
\caption{The same as in Table~\ref{tab:5} for the $\mathrm{Ho}^{66+} - \bar{p}$ compound. For this quasi-molecule, the values of the Zeeman shift in the $3$, $4$, $5$, $10$ and $100$ T fields are indicated in the last five columns, respectively.}
\label{tab:7}
\begin{tabular}{ c c c c c c c }
    \hline
    \hline
        \multirow{2}{*}{$R$, fm} & \multicolumn{1}{c|}{\multirow{2}{*}{$E_{\mathrm{ADKB}}$, a.u.}} & \multicolumn{5}{c}{$ | \Delta E_{\mathrm{Z}} | \times 10^6 $} \\ \cline{3-7} 
                       & \multicolumn{1}{c|}{}                                  & $B = 3$ T  & $B = 4$ T & $B = 5$ T & $B = 10$ T & $B = 100$ T \\ \hline
        $0.0$ & $-2321.02291453$ & $5.86$ & $7.81$ & $9.76$ & $19.52$ & $195.20$ \\ 
        $157.96336$ & $-2323.82451644$ & $5.85$ & $7.81$ & $9.76$ & $19.51$ & $195.14$ \\ 
        $315.92672$ & $-2329.01021529$ & $5.85$ & $7.80$ & $9.75$ & $19.50$ & $195.04$ \\ 
        $473.89007$ & $-2334.84243831$ & $5.85$ & $7.80$ & $9.75$ & $19.49$ & $194.95$ \\ 
        $631.85343$ & $-2340.60253542$ & $5.85$ & $7.79$ & $9.74$ & $19.49$ & $194.87$ \\ 
        $789.81679$ & $-2345.96445844$ & $5.84$ & $7.79$ & $9.74$ & $19.48$ & $194.80$ \\ 
        $947.78015$ & $-2350.80386659$ & $5.84$ & $7.79$ & $9.74$ & $19.47$ & $194.75$ \\ 
        $1105.74350$ & $-2355.10160294$ & $5.84$ & $7.79$ & $9.74$ & $19.47$ & $194.71$ \\ 
        $1263.70686$ & $-2358.87442259$ & $5.84$ & $7.79$ & $9.73$ & $19.47$ & $194.68$ \\ 
        $1421.67022$ & $-2362.16494487$ & $5.84$ & $7.79$ & $9.73$ & $19.47$ & $194.66$ \\ 
        $1579.63358$ & $-2365.03189436$ & $5.84$ & $7.79$ & $9.73$ & $19.46$ & $194.65$ \\ 
        $1737.59694$ & $-2367.53574608$ & $5.84$ & $7.78$ & $9.73$ & $19.46$ & $194.64$ \\ 
        $1895.56029$ & $-2369.72982539$ & $5.84$ & $7.78$ & $9.73$ & $19.46$ & $194.63$ \\ 
        $2053.52365$ & $-2371.65927895$ & $5.84$ & $7.78$ & $9.73$ & $19.46$ & $194.63$ \\ 
        $2211.48701$ & $-2373.36217338$ & $5.84$ & $7.78$ & $9.73$ & $19.46$ & $194.63$ \\ 
        $2369.45037$ & $-2374.87051899$ & $5.84$ & $7.78$ & $9.73$ & $19.46$ & $194.62$ \\ 
        $2527.41373$ & $-2376.21115360$ & $5.84$ & $7.78$ & $9.73$ & $19.46$ & $194.62$ \\ 
        $2685.37708$ & $-2377.40656969$ & $5.84$ & $7.78$ & $9.73$ & $19.46$ & $194.62$ \\ 
        $2843.34044$ & $-2378.47569427$ & $5.84$ & $7.78$ & $9.73$ & $19.46$ & $194.62$ \\ 
        $3001.30380$ & $-2379.43460173$ & $5.84$ & $7.78$ & $9.73$ & $19.46$ & $194.62$ \\ 
        $3159.26716$ & $-2380.29712949$ & $5.84$ & $7.79$ & $9.73$ & $19.46$ & $194.62$ \\ 
    \hline
    \hline
    \end{tabular}
\end{table}
%}
\begin{table}[ht]
\centering
\caption{The same as in Table~\ref{tab:7} for the $\mathrm{Re}^{74+} - \bar{p}$ compound. The values of the Zeeman shift in the $2$, $3$, $4$, $5$, $10$ and $100$ T fields are indicated in the last six columns, respectively.}
\label{tab:8}
\begin{tabular}{ c c c c c c c c }
    \hline
    \hline
        \multirow{2}{*}{$R$, fm} & \multicolumn{1}{c|}{\multirow{2}{*}{$E_{\mathrm{ADKB}}$, a.u.}} & \multicolumn{6}{c}{$ | \Delta E_{\mathrm{Z}} | \times 10^6 $}            \\ \cline{3-8} 
                       & \multicolumn{1}{c|}{}                                           & $B = 2$ T & $B = 3$ T & $B = 4$ T & $B = 5$ T & $B = 10$ T & $B = 100$ T \\ \hline
        $0.0$ & $-2972.30805239$ & $3.81$ & $5.71$ & $7.61$ & $9.51$ & $19.03$ & $190.29$ \\ 
        $141.11393$ & $-2976.09312938$ & $3.80$ & $5.71$ & $7.61$ & $9.51$ & $19.02$ & $190.21$ \\ 
        $282.22787$ & $-2982.65931510$ & $3.80$ & $5.70$ & $7.60$ & $9.50$ & $19.01$ & $190.10$ \\ 
        $423.34180$ & $-2989.80282984$ & $3.80$ & $5.70$ & $7.60$ & $9.50$ & $19.00$ & $189.98$ \\ 
        $564.45573$ & $-2996.70635801$ & $3.80$ & $5.70$ & $7.59$ & $9.49$ & $18.99$ & $189.89$ \\ 
        $705.56966$ & $-3003.04369779$ & $3.80$ & $5.69$ & $7.59$ & $9.49$ & $18.98$ & $189.81$ \\ 
        $846.68360$ & $-3008.68956528$ & $3.79$ & $5.69$ & $7.59$ & $9.49$ & $18.97$ & $189.76$ \\ 
        $987.79753$ & $-3013.65734919$ & $3.79$ & $5.69$ & $7.59$ & $9.49$ & $18.97$ & $189.71$ \\ 
        $1128.91146$ & $-3018.00053903$ & $3.79$ & $5.69$ & $7.59$ & $9.48$ & $18.97$ & $189.68$ \\ 
        $1270.02540$ & $-3021.77354569$ & $3.79$ & $5.69$ & $7.59$ & $9.48$ & $18.97$ & $189.66$ \\ 
        $1411.13933$ & $-3025.03933807$ & $3.79$ & $5.69$ & $7.59$ & $9.48$ & $18.96$ & $189.65$ \\ 
        $1552.25326$ & $-3027.86970304$ & $3.79$ & $5.69$ & $7.58$ & $9.48$ & $18.96$ & $189.64$ \\ 
        $1693.36720$ & $-3030.33588984$ & $3.79$ & $5.69$ & $7.58$ & $9.48$ & $18.96$ & $189.63$ \\ 
        $1834.48113$ & $-3032.49857997$ & $3.79$ & $5.69$ & $7.58$ & $9.48$ & $18.96$ & $189.62$ \\ 
        $1975.59506$ & $-3034.40616085$ & $3.79$ & $5.69$ & $7.58$ & $9.48$ & $18.96$ & $189.62$ \\ 
        $2116.70900$ & $-3036.09723779$ & $3.79$ & $5.69$ & $7.58$ & $9.48$ & $18.96$ & $189.62$ \\ 
        $2257.82293$ & $-3037.60300317$ & $3.79$ & $5.69$ & $7.58$ & $9.48$ & $18.96$ & $189.62$ \\ 
        $2398.93686$ & $-3038.94886299$ & $3.79$ & $5.69$ & $7.58$ & $9.48$ & $18.96$ & $189.62$ \\ 
        $2540.05079$ & $-3040.15561730$ & $3.79$ & $5.69$ & $7.58$ & $9.48$ & $18.96$ & $189.61$ \\ 
        $2681.16473$ & $-3041.24039834$ & $3.79$ & $5.69$ & $7.59$ & $9.48$ & $18.96$ & $189.62$ \\ 
        $2822.27866$ & $-3042.21745820$ & $3.79$ & $5.69$ & $7.58$ & $9.48$ & $18.96$ & $189.61$ \\ 
    \hline
    \hline
    \end{tabular}
\end{table}
\begin{table}[ht]
\centering
\caption{The same as in Table~\ref{tab:7} for the $\mathrm{U}^{91+} - \bar{p}$ compound. The Zeeman shift is presented at the field strength $1$, $2$, $3$, $4$, $5$, $10$, $100$ T.}
\label{tab:9}
\begin{tabular}{ c c c c c c c c c }
    \hline
    \hline
        \multirow{2}{*}{$R$, fm} & \multicolumn{1}{c|}{\multirow{2}{*}{$E_{\mathrm{ADKB}}$, a.u.}} & \multicolumn{7}{c}{$ | \Delta E_{\mathrm{Z}} | \times 10^6 $}                                   \\ \cline{3-9} 
                       & \multicolumn{1}{c|}{}                                           & $B = 1$ T            & $B = 2$ T & $B = 3$ T & $B = 4$ T & $B = 5$ T & $B = 10$ T & $B = 100$ T \\ \hline
        $0.0$ & $-4731.61940826$ & $1.77$ & $3.54$ & $5.31$ & $7.08$ & $8.85$ & $17.71$ & $177.06$ \\ 
        $115.03853$ & $-4739.21337367$ & $1.77$ & $3.54$ & $5.31$ & $7.08$ & $8.85$ & $17.69$ & $176.92$ \\ 
        $230.07706$ & $-4750.31646276$ & $1.77$ & $3.53$ & $5.30$ & $7.07$ & $8.84$ & $17.67$ & $176.74$ \\ 
        $345.11560$ & $-4761.37516295$ & $1.76$ & $3.53$ & $5.30$ & $7.06$ & $8.83$ & $17.66$ & $176.58$ \\ 
        $460.15413$ & $-4771.48208068$ & $1.76$ & $3.53$ & $5.29$ & $7.06$ & $8.82$ & $17.64$ & $176.46$ \\ 
        $575.19266$ & $-4780.37762099$ & $1.76$ & $3.53$ & $5.29$ & $7.05$ & $8.82$ & $17.64$ & $176.36$ \\ 
        $690.23119$ & $-4788.10088334$ & $1.76$ & $3.53$ & $5.29$ & $7.05$ & $8.82$ & $17.63$ & $176.30$ \\ 
        $805.26973$ & $-4794.73586551$ & $1.76$ & $3.52$ & $5.29$ & $7.05$ & $8.81$ & $17.62$ & $176.25$ \\ 
        $920.30826$ & $-4800.37950112$ & $1.76$ & $3.52$ & $5.29$ & $7.05$ & $8.81$ & $17.62$ & $176.21$ \\ 
        $1035.34679$ & $-4805.23045976$ & $1.76$ & $3.52$ & $5.29$ & $7.05$ & $8.81$ & $17.62$ & $176.19$ \\ 
        $1150.38532$ & $-4809.43838298$ & $1.76$ & $3.52$ & $5.28$ & $7.05$ & $8.81$ & $17.62$ & $176.17$ \\ 
        $1265.42386$ & $-4813.08395527$ & $1.76$ & $3.52$ & $5.28$ & $7.05$ & $8.81$ & $17.61$ & $176.16$ \\ 
        $1380.46239$ & $-4816.22345098$ & $1.76$ & $3.52$ & $5.28$ & $7.05$ & $8.81$ & $17.61$ & $176.15$ \\ 
        $1495.50092$ & $-4818.91377728$ & $1.76$ & $3.52$ & $5.28$ & $7.05$ & $8.81$ & $17.61$ & $176.15$ \\ 
        $1610.53945$ & $-4821.22305291$ & $1.76$ & $3.52$ & $5.28$ & $7.05$ & $8.81$ & $17.61$ & $176.15$ \\ 
        $1725.57799$ & $-4823.22648989$ & $1.76$ & $3.52$ & $5.28$ & $7.05$ & $8.81$ & $17.61$ & $176.15$ \\ 
        $1840.61652$ & $-4824.99169849$ & $1.76$ & $3.52$ & $5.28$ & $7.05$ & $8.81$ & $17.61$ & $176.15$ \\ 
        $1955.65505$ & $-4826.56937959$ & $1.76$ & $3.52$ & $5.28$ & $7.05$ & $8.81$ & $17.62$ & $176.15$ \\ 
        $2070.69358$ & $-4827.99460806$ & $1.76$ & $3.52$ & $5.28$ & $7.05$ & $8.81$ & $17.61$ & $176.15$ \\ 
        $2185.73211$ & $-4829.29165902$ & $1.76$ & $3.52$ & $5.28$ & $7.04$ & $8.81$ & $17.61$ & $176.15$ \\ 
        $2300.77065$ & $-4830.47781493$ & $1.76$ & $3.52$ & $5.28$ & $7.04$ & $8.81$ & $17.61$ & $176.15$ \\ 
    \hline
    \hline
    \end{tabular}
\end{table}

\end{document}